\documentclass[aps,pra,reprint,superscriptaddress,nofootinbib]{revtex4-2}
\usepackage{graphicx}% Include figure files
\usepackage{mathpazo}
\usepackage{dcolumn}% Align table columns on decimal point
\usepackage{xcolor}
\usepackage{bm}
\usepackage{tcolorbox}
\usepackage{braket}
\usepackage{lipsum}
\usepackage{amsmath, amsthm, amsfonts, amssymb, bbm}
\usepackage{mathtools}
\usepackage{pifont}
\usepackage[colorlinks,linkcolor=blue,urlcolor=blue, citecolor=blue]{hyperref}% add hypertext capabilities

\newcommand{\ketbra}[2]{|#1\rangle\! \langle #2|}

\newtheorem{theorem}{Theorem}[section]
\newtheorem{corollary}{Corollary}[section]

\newcommand{\rvline}{\hspace*{-\arraycolsep}\vline\hspace*{-\arraycolsep}}
\DeclareMathAlphabet{\mymathbb}{U}{BOONDOX-ds}{m}{n}

\newcommand\scalemath[2]{\scalebox{#1}{\mbox{\ensuremath{\displaystyle #2}}}}

\begin{document}

\preprint{APS/123-QED}

\title{Quantum Coding with Finite Thermodynamic Resources}

\author{Jake Xuereb}
\affiliation{Vienna Center for Quantum Science and Technology, Atominstitut, TU Wien, 1020 Vienna, Austria}
%\email{jake.xuereb@tuwien.ac.at}
\author{Tiago Debarba}
\affiliation{Vienna Center for Quantum Science and Technology, Atominstitut, TU Wien, 1020 Vienna, Austria}
\affiliation{Departamento Acad{\^ e}mico de Ci{\^ e}ncias da Natureza, Universidade Tecnol{\'o}gica Federal do Paran{\'a} (UTFPR), Campus Corn{\'e}lio Proc{\'o}pio, Avenida Alberto Carazzai 1640, Corn{\'e}lio Proc{\'o}pio, Paran{\'a} 86300-000, Brazil.}
\author{Marcus Huber}
\affiliation{Vienna Center for Quantum Science and Technology, Atominstitut, TU Wien, 1020 Vienna, Austria}
\affiliation{Institute for Quantum Optics and Quantum Information (IQOQI), Austrian Academy of Sciences,
Boltzmanngasse 3, 1090 Vienna, Austria}
\author{Paul Erker}
\affiliation{Vienna Center for Quantum Science and Technology, Atominstitut, TU Wien, 1020 Vienna, Austria}
\affiliation{Institute for Quantum Optics and Quantum Information (IQOQI), Austrian Academy of Sciences,
Boltzmanngasse 3, 1090 Vienna, Austria}

\date{\today}

\begin{abstract}Quantum direct coding or Schumacher compression generalised the ideas of Shannon theory, gave an operational meaning to the von Neumann entropy and established the term \textit{qubit}. But remembering that information processing is carried out by physical processes prompts one to wonder what thermodynamic resources are required to compress quantum information and how they constrain one's ability to perform this task. That is, if Alice and Bob only have access to thermal quantum states and clocks with finite accuracy, how well can they measure, encode and decode pure quantum state messages? In this work we examine these questions by modelling Alice's typical measurement as a unitary involving a measurement probe, investigating imperfect timekeeping on encoding and decoding and considering the role of temperature in Bob's appended qubits. In doing so, we derive fidelity bounds for this protocol involving the correlations Alice can form with their measurement probe, the variance of the clock's ticks and the temperature of Bob's qubits. Finally, we give an insight into the entropy produced by these two agents throughout the compression protocol by relating the resources they use to a quantum thermodynamic cooling protocol. 
\end{abstract}

\maketitle

\section{Introduction}
%%What is compression?%%
The weak law of large numbers (W.L.L.N)~\cite{cover} ensures that the entropy of large sequences of samples of an independent and identically distributed (i.i.d.) source converges to the entropy of the source itself as the number of samples $n$ grows. That is, large sequences of samples converge to \textit{typical} sequences-- sequences whose sample entropy is very close to the Shannon entropy of the source. Shannon made use of this idea together with the observation that the set of typical sequences of a given length has smaller cardinality than the larger set of possible sequences of this same length, to come up with the seminal noiseless coding theorem~\cite{shannon} marking the birth of the field of information theory. Schumacher compression~\cite{schumacher} is the quantum analogue of this idea where typical sequences of pure quantum states sampled from an i.i.d.~quantum source are block encoded into the typical subspace of the sample space. This allows one to communicate quantum information in a compressed form at a rate given by the von Neumann entropy of the source.

%%The difference between quantum information theory and information theory%%
Classical and quantum compression are not just different from a mathematical perspective but also from a physical point of view, Shannon and Schumacher compression vary operationally at two points. Firstly, in order to compress a sequence of quantum states, it must be measured to determine that it is typical w.r.t.~the source, something which is classically unnecessary and can alter the quantum message if it is not sufficiently long~\cite{winter}. Secondly, whilst the encoding in Shannon compression is a Boolean function, Schumacher compression incorporates such an encoding function into a unitary operation generated by a physical process. Both these differences present opportunities for error in quantum coding schemes to creep in due to access to only finite resources. 

\begin{table}[b]
\begin{tabular}{|l|ll|}
\hline
\multicolumn{1}{|c|}{Source} & \multicolumn{2}{l|}{Impact on Fidelity}                                 \\ \hline
Alice's Thermal Probe        & \multicolumn{1}{l|}{Thm.\ref{thm:alice}}  & Hyperb. in $\beta$ + Stat. Mixing \\ \hline
Timekeeping                  & \multicolumn{1}{l|}{Eq.\ref{eq:timing}}   & Exp. in $\sigma^2$                \\ \hline
Bob's Thermal Qubits         & \multicolumn{1}{l|}{Eq.\ref{eq:bob_fid}} & Hyperb. in $\beta$, Exp. in $J$   \\ \hline
\end{tabular}
\caption{Table summarising main results showing relationships between fidelity and thermodynamic resource, where $\beta = 1/k_B T$ is the inverse temperature of Alice's qubit measurement probe or Bob's appended qubits, $\sigma$ is the variance in the clock's ticks and $J$ is the number of thermal qubits appended by Bob.}
\end{table}
%%Discuss previous connection to rate distortion and the motivation to examine information theory with access to finite resources.%%
Previous work~\cite{hsieh_2021,korzekwa_22} has investigated thermodynamic resources required to encode and communicate classical information quantumly but the thermodynamic resources required to compress quantum information has thus far not been examined. Contemporary tools from quantum thermodynamics~\cite{thermo_review,Guryanova_2020,xuereb2023} now empower us to relate these abstract errors that can arise in quantum compression scenario to at least two physical origins, temperature and timekeeping. 

\begin{figure*}[t]
    \centering
\includegraphics[width = 0.78\textwidth]{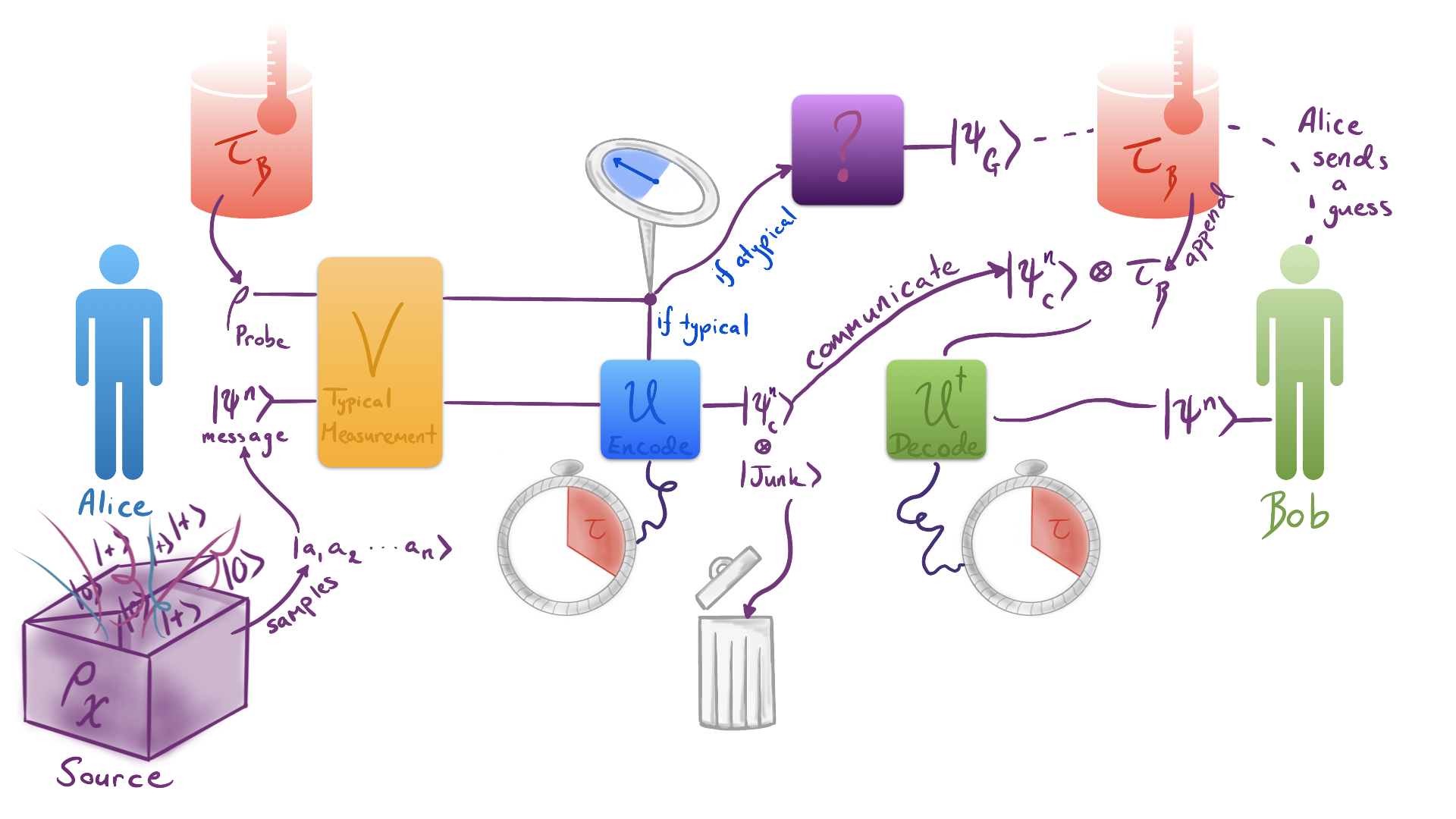}
    \caption{An illustration of the setup for \textit{thermodynamically realistic quantum coding} considered in this work where Alice and Bob have access to imperfect clocks with a finite accuracy which they can use to generate unitaries for encoding and decoding, and qubit thermal states to be used as measurement probes or to append for decoding.}
\label{fig:illustration}
\end{figure*}

In this work we investigate the physical resources required to compress i.i.d.~quantum information by considering what we call \textit{thermodynamically realistic quantum coding} illustrated in Fig. \ref{fig:illustration}, where the agents carrying out this protocol have access to thermal qubit states and clocks with a finite accuracy as resources. First in Sec.~\ref{sec:typical_measurement}, inspired by recent work examining the thermodynamic resources required to carry out projective measurements~\cite{Guryanova_2020} we model the typical measurement as a unitary process with an ancillary measurement probe. In doing so we derive a bound which shows that the quality of measurement probe an agent has access to limits the average protocol fidelity they are able to achieve. In particular, an agent may only recover the conventional average fidelity bound~\cite{Preskill1998} for quantum coding with access to at least one of 3 diverging resources; dimension, temperature or energy gap. In Sec. \ref{sec:timing}, we consider imperfectly timed encoding and decoding unitaries~\cite{xuereb2023} where the protocol is dephased in the energy eigenbasis of the encoding Hamiltonian resulting in an exponential impact on the achievable fidelity in terms of the accuracy of the clocks the agents have access to. In Sec.~\ref{sec:decoding}, we consider the impact of decoding with access to only thermal qubits and derive a fidelity relationship that is exponential in the number of qubits Alice erases in the compression step. Finally, in Sec.~\ref{sec:entropy_production}, we consider that Alice and Bob can improve their fidelities by cooling their qubits using a quantum thermodynamic protocol, at the cost of producing entropy. This allows us to unify our insights into one statement--\textit{perfect quantum coding is not possible with finite entropy production}\footnote{Entropy production~\cite{landi_paternostro} or irreversible work $\Sigma$ is directly related to the heat dissipated to a thermal environment in a process by $\Sigma~=~\Delta S + \beta \Delta Q$. As such these terms are at times used interchangeably in the text.}.
 
\subsection{Quantum coding}%%Information Theoretic background
In Schumacher compression~\cite{schumacher,Preskill1998,wilde_2013} an agent has access to an i.i.d.~quantum information source $\chi$ with $l$ pure quantum state letters in an alphabet $\Sigma = \left\{\ket{\phi_i}\right\}^{l}_{i=1}$ associated with probabilities $p_i$ of the agent sampling the letter $\ket{\phi_i}$. Without loss of generality, we will assume these letters to be qubits in this work. A message $\ket{\psi^n}$ of $n$ qubit samples from this i.i.d.~quantum source will be distributed by sample according to $\rho_\chi~=~\sum^{l}_{i} p_i \ketbra{\phi_i}{\phi_i}$ such that the probability of sampling a string of pure states $\ket{\psi^n} = \ket{a_1} \otimes \ket{a_2} \dots \otimes \ket{a_n}$ with each $a_k \in \Sigma$ is given by $p(\psi^n) = p(a_1a_2\dots a_n) = p(a_1)p(a_2)\dots p(a_n)$. Consider an agent, Alice who obtains $\ket{\psi^n}$ by sampling and now wishes to send this message to another agent Bob using the least qubits possible by focusing on likely messages. Since the source is i.i.d.~each string is distributed sample by sample according to $\rho_\chi = \sum^{d}_{j=1}\gamma_j \ketbra{j}{j}$ by the spectral decomposition of $\rho_\chi$ and so messages with large overlap with $\ket{j}^{\otimes n}$ corresponding to the largest $\gamma_j$ will appear most frequently. This notion is known as \textit{typicality} in information theory. Formalising this, each eigenstate of $\rho_\chi^{\otimes n}$ is a possible message and the weight of the eigenstate in $\rho_\chi^{\otimes n}$, the likelihood of sampling the string. For qubits these eigenstates are binomially distributed according to their eigenvalues $\lambda_i = \binom{n}{i}\gamma^{n-i}_1\gamma^{i}_2$ where degeneracies have been summed over, meaning that some products of eigenvalues have higher weight resulting in a high weight eigenstate i.e. a highly likely message. Following this reasoning \cite{schumacher} states that a string of eigenvalues corresponding to a state, is said to be $\epsilon$-typical if it satisfies $|-\frac{1}{n}\sum^{2^n}_{i=1} \log \lambda_i - S(\rho_\chi)| \leq \epsilon$. In this way, the projective Hilbert space of $n$ qubits can be partitioned into two subspaces corresponding to the states whose eigenvalues form $\epsilon$-typical sequences and those that do not. The W.L.L.N. ensures that when an agent samples long enough from an i.i.d.~source, the sample entropy converges to the von Neumann entropy of the source. That is, all large enough messages are typical or mathematically given $\epsilon, \delta > 0$ there exists $N \in \mathbb{N} \, : \forall n \geq N$ \begin{align}
     \mathbb{P}\left[\left|-\frac{1}{n}\sum^{2^n}_{i=1} \log \lambda_i - S(\rho_\chi)\right| > \epsilon \right] < \delta,
\end{align}
where $\mathbb{P}[\cdot]$ denotes probability~\cite{schumacher}. Together with the realisation that the set of all $n$ sample messages from $\chi$ has cardinality $2^{n \log l}$ and the set of $\epsilon$-typical sequences has cardinality upper bounded by $2^{n(S(\rho_\chi) + \epsilon)}$~\cite{Preskill1998, wilde_2013}, this allows Alice to block encode an $n$ qubit message $\ket{\psi^n}$ into a compressed $n(S(\rho_\chi) + \epsilon)$ qubit message. 
%%Protocol, Typical Measurement, Typical Projector, Unitary Encoding and Decoding, Guess State, Fidelity.

Since $\ket{\psi^n}$ is a quantum message, before block encoding Alice must determine by measurement that their message is typical with respect to the source $\rho_\chi$ and so lies prominently along the typical subspace defined by the $\epsilon$-typical subspace projector $\Lambda^\epsilon_n~=~\sum_{\lambda^n_k \in T^\epsilon_n} \ketbra{\psi^n_k}{\psi^n_k}$ comprised of the projectors of states $\ket{\psi^n_k}$ with eigenvalues $\lambda^n_k$ that are in the $\epsilon$-typical. The typical measurement can be described by the channel
\begin{gather}\mathcal{T}^{\epsilon}_n(\sigma) = \Lambda^\epsilon_n\sigma\Lambda^\epsilon_n + (\mathbbm{1}_n - \Lambda^\epsilon_n)\sigma (\mathbbm{1}_n - \Lambda^\epsilon_n)
\label{eq:typ_meas}
\end{gather}
and is a binary measurement with two outcomes, typical or atypical. One might be concerned that this measurement could spoil the content of the quantum message, but Winter showed that the typical projector is a gentle operator~\cite{winter}. That is, if the message is sufficiently large its overlap with the typical subspace is large $\bra{\psi^n}\Lambda^\epsilon_n\ket{\psi^n} = 1 - \delta \, : \, 0 < \delta \leq 1$  and so the state of the message is only slightly perturbed $||\ketbra{\psi^n}{\psi^n} - \Lambda^\epsilon_n \ketbra{\psi^n}{\psi^n} \Lambda^\epsilon_n||_1\leq 2\sqrt{\delta}$ by typical measurement, where the perturbation is smaller the larger the message is. 

If Alice determines that their message state is typical, they encode by applying an $n$ qubit unitary $U_f$ which is a coherent implementation of the classical function $f~:~\psi^n_k\, \to~\, \{0,1\}^{n(S(\rho_\chi)+\epsilon)}, \, \psi^n_k \, \in \, T^\epsilon_n$ which effectively rotates the message state such that the typical subspace is concentrated on the first $\lceil n S(\rho_\chi) + \epsilon\rceil$ qubits and the remainder are traced out, compressing the message to $\ket{\psi^n_c}$ which is communicated to Bob. Bob then appends the received compressed message with $(n - \lceil n S(\rho_\chi) + \epsilon\rceil) = J$ qubits in the $\ket{0}$ state to decode by applying $U^\dagger_f$. In the case that Alice obtains an atypical message, they then send a guess state $\ket{\psi_G}$ to Bob which can be e.g. the eigenket $\ket{j}$ with largest $\lambda_j$. At the end of this protocol, Bob will have received a statistical mixture of compressed and guess states leading to the ensemble
\begin{gather}
    \rho_R = \Lambda^\epsilon_n\ketbra{\psi^n}{\psi^n}\Lambda^\epsilon_n + \bra{\psi^n}\left(\mathbbm{1} - \Lambda^\epsilon_n \right)\ket{\psi^n} \ketbra{\psi_G}{\psi_G},
\end{gather}
where the first contribution is due to the decoded states and the second due to the guess states. Resulting in a fidelity between the message sent $\ket{\psi^n}$ and the received ensemble $\rho_R$ of
\begin{align}
    \mathcal{F}\left(\ket{\psi^n}, \rho_R\right) = |\bra{\psi^n}&\Lambda^\epsilon_n\ket{\psi^n}|^2  \\ 
    &+  \bra{\psi^n}\left(\mathbbm{1} - \Lambda^\epsilon_n \right)\ket{\psi^n}|\braket{\psi_G|\psi}|^2, \nonumber 
\end{align}
where the fidelity between an arbitrary pure state $\ket{\mu}$ and a mixed state $\sigma$ can be expressed as $\mathcal{F}(\sigma, \ket{\mu})~=~\bra{\mu}\sigma\ket{\mu}$~\cite{Preskill1998}. If $\mathcal{F}$ is averaged over the set of possible messages from the source $\rho_\chi$ it can be shown to be lower bounded~\cite{schumacher,Preskill1998} by
\begin{gather}
    \overline{\mathcal{F}} = \sum^{n^l}_{j=1}p(\psi^n_k)\mathcal{F}(\ket{\psi^n_k},\rho_R) \geq 1 - 2 \delta.
\end{gather}
where $p(\psi^n_k)$ is the probability of a given message state. 

An illustrative example of Schumacher compression for a two letter qubit alphabet and qubit messages of 3 samples is given in Appendix~\ref{appendix:instructive} together with further explanation of the theory.
\subsection{Thermodynamically realistic quantum coding} 
%Introduce the setup we will deal with 
The two operations we will investigate within this work that require thermodynamic resources are typical measurement and encoding--decoding. Quantum measurement can be thought of as the task of correlating a system of interest with the state of an ancillary system called the measurement probe such that these correlations encode the measurement statistics~\cite{von2013mathematische,zurek,petruccione,Paris_2012}. 
The amount and quality of correlations formed with the measurement probe are limited by the purity and dimension of the probe. Indeed, it has been shown that perfect projective measurements require a diverging resources~\cite{Guryanova_2020} (dimension, temperature or energy gap) to create perfect correlations between the probe and system of interest. On the other hand, encoding and decoding are unitary gates generated through a timed physical process described by some Hamiltonian. Accurate and precise timekeeping comes at the cost of entropy production~\cite{thermo_clock} and imperfect timekeeping leads to dephasing error~\cite{xuereb2023}. Beyond unitary operations, decoding also requires Bob to append $J$ qubits in the $\ket{0}$ state, the preparation of which also requires diverging resources~\cite{Taranto_2023}. 

This begs the question, \textit{if Alice and Bob are only allowed access to thermal states of a Hamiltonian $H_d$ with dimension $d$ at inverse temperature $\beta$ and clocks that tick with a finite variance $\sigma$-- with what average fidelity can they compress and recover i.i.d.~quantum information?} In our scheme we attempt to answer this question by allowing our agents to produce a finite amount of dissipation in a protocol where Alice carries out the typical measurement with access to a thermal probe, both agents together carry out imperfectly timed unitaries for encoding and decoding and Bob decodes by appending the thermal qubits they have access to. Later on in Sec.\ref{sec:entropy_production} we also allow our agents to create better resources out of those they have access to, at expense of producing entropy in a cooling protocol.
\section{Typical Measurement}
\label{sec:typical_measurement}
Since the typical measurement is binary, Alice needs to correlate their message with at least a qubit to capture the statistics of the two outcomes. To achieve this we dilate $\mathcal{T}^\epsilon_n$ using an ancillary qubit $\tau_\beta = e^{-\beta H}/\text{tr}\{e^{-\beta H}\}$ in a thermal state of $H_A = -\hbar \omega \sigma_z$ at inverse temperature $\beta = 1/k_B T$ where $\omega$ is the energy splitting of the qubit which we will fix throughout this work for the qubits involved in the compression protocol. A global interaction $V$ correlates the message state and the probe qubit to give 
\begin{align}
    \mathcal{T}^\epsilon_n(\ketbra{\psi^n}{\psi^n}) = \text{tr}_A\left\{V(\ketbra{\psi^n}{\psi^n} \otimes \tau_\beta)V^\dagger \right\}
\end{align}
and recover the correct statistics. The choice of unitary $V = \Lambda^\epsilon_n \otimes \mathbbm{1}_1 + \left(\mathbbm{1}_n - \Lambda^\epsilon_n\right) \otimes \sigma_x$ is a form of controlled-not on the probe qubit which is conditioned on whether the message is typical or not. This induces correlations that encode the expected outcomes of the typical measurement correctly if the qubit probe is in the $\ket{0}$ state i.e at zero temperature or has an infinitely gapped Hamiltonian as shown in \cite{Guryanova_2020}. If the temperature is non-zero then the populations of the thermal qubit will bias the measurement outcomes in a way which does not reflect the content of the message Alice is trying to compress. That is, when Alice's probe reads 0 they do not obtain $\rho_{\psi^n | 0} = \Lambda^\epsilon_n\ketbra{\psi^n}{\psi^n}\Lambda^\epsilon_n$ as expected but rather \begin{align}\rho_{\psi^n | 0} = \frac{e^{-\beta \hbar \omega}}{\mathcal{Z}}\Lambda^\epsilon_n&\ketbra{\psi^n}{\psi^n}\Lambda^\epsilon_n \\
&+ \frac{e^{\beta \hbar \omega}}{\mathcal{Z}}(\mathbbm{1}_n - \Lambda^\epsilon_n)\ketbra{\psi^n}{\psi^n}(\mathbbm{1}_n-\Lambda^\epsilon_n) \nonumber\end{align} where $\mathcal{Z} = \text{tr}\{e^{-\beta H_A}\}.$ 

To mitigate this with finite resources, let us encode the binary outcomes into the populations of a qudit in a thermal state $\tau_{\beta,d}$ of a Hamiltonian $H_d = \sum^{d}_{i=1} E_i \ketbra{i}{i}$ at inverse temperature $\beta$. This allows us to correlate the outcome of the message being typical with the qudit occupying the first $d/2$ energy levels using $V_d = \Lambda^\epsilon_n \otimes \mathbbm{1}_d + \left(\mathbbm{1}_n - \Lambda^\epsilon_n\right) \otimes \sigma_x^{\otimes \, d/2}$, resulting in the typical outcome is associated with the post-measurement state \begin{align}\rho_{\psi^n | 0} = \sum^{d/2}_{i = 1}&\frac{e^{-\beta E_i}}{\mathcal{Z}}\Lambda^\epsilon_n\ketbra{\psi^n}{\psi^n}\Lambda^\epsilon_n \\
+& \sum^{d}_{j = d/2+1}\frac{e^{-\beta E_j}}{\mathcal{Z}}(\mathbbm{1}_n - \Lambda^\epsilon_n)\ketbra{\psi^n}{\psi^n}(\mathbbm{1}_n-\Lambda^\epsilon_n) \nonumber
\end{align}
where increasing the dimension allows one to approach the correct measurement statistics even with finite resources. The quantity $C_{\text{Max}}^{(\beta, H_d,d)}$ defined in \cite{Guryanova_2020} is a function related to the maximum correlations which can be formed between the system of interest and the measurement probe depending on the temperature of the probe, its dimension and the spectral gap of its Hamiltonian. In our case 
\begin{equation}
    C_{\text{Max}}^{(\beta, H_d,d)} = \sum^{d/2}_{i = 1}\frac{e^{-\beta E_i}}{\mathcal{Z}} \in \left[\frac{1}{2},1\right],
\end{equation}
and if this quantity is equal to 1 through the divergence of $\beta$, the gap $H_d$ or $d$ we have perfect measurement.

When Alice makes a typical measurement of their message $\ket{\psi^n}$ using a thermal probe $\tau_{\beta, d}$ the bias due to the populations means that sometimes atypical messages are sent to Bob when a guess state should have been sent and vice-versa, confusing what is typical and atypical. As a result of this statistical mixing Bob receives the ensemble \begin{align}
\widetilde{\rho_R} & = C\Lambda^\epsilon_n\ketbra{\psi^n}{\psi^n}\Lambda^\epsilon_n+ \left(1 - C\right){\Lambda^\epsilon_n}^\perp\ketbra{\psi^n}{\psi^n}{\Lambda^\epsilon_n}^\perp  \\ &+ \left(\left(1 - C\right)\bra{\psi^n}\Lambda^\epsilon_n\ket{\psi^n}
+C\bra{\psi^n}{\Lambda^\epsilon_n}^\perp\ket{\psi^n}\right)\ketbra{\psi_G}{\psi_G} \nonumber
 \end{align}
 where for convenience we have used the notation $C_{\text{Max}}^{(\beta, H_A,d)} = C$ and $\mathbbm{1}_n - \Lambda^\epsilon_n = {\Lambda^\epsilon_n}^\perp$. We can characterise the impact that this finite resource has on an agent's ability to compress i.i.d.~quantum information by considering the fidelity of the compressed information with the original information across every possible $n$ sample message one could sample from the source.
\begin{theorem}
\label{thm:alice}
For an i.i.d.~quantum source $\chi$ with density $\rho_\chi~=~\sum^{l}_{i=1} p_i \ketbra{\phi_i}{\phi_i}$ per sample, the average fidelity $\overline{\mathcal{F}}$ over possible $n$ sample messages $\ket{\psi^n}$ sampled from $\chi$ of a quantum coding protocol where the sender Alice has access to a measurement probe in a thermal state $\tau_{\beta,d}$, is lower bounded by a function of the correlations of the measurement probe $C_{\mathrm{Max}}^{(\beta, H_A,d)}$ as
\begin{align}
  \overline{\mathcal{F}}  \geq  C_{\mathrm{Max}}^{(\beta, H_A,d)}(1 - 2\delta) + (1 - C_{\mathrm{Max}}^{(\beta, H_A,d)})\delta^2, \nonumber
\end{align}such that the expected bound $1-2\delta$ is only recovered when the probe has one of three diverging properties.
\end{theorem}
\emph{Here is a sketch of the proof}. We consider the average fidelity $\overline{\mathcal{F}} = \sum^{n^l}_{j = 1}p(\psi^n_j)\bra{\psi^n_j}\widetilde{\rho_R}\ket{\psi^n_j}$ between the mixed ensemble state Bob receives and the $n$ sample pure state message sampled by Alice, averaging over the set of possible messages denoted by $\{\ket{\psi^n_j}\}^{n^l}_{j=1}$ where each message occurs with probability $p(\psi^n_j)$. This sum will have two main contributions, one from ensembles which were compressed and one from ensembles that led to a guess state being sent. The compression contribution can be split into terms which lie along the typical subspace and those that do not, which result from the bias from the non-zero temperature thermal probe. The typical compression contribution is formed of squared terms close to 1 allowing them to be lower bounded by $x^2 \geq 2(x - 1)$. This results in a sum of terms equal to the expectation of the $n$ sample density with the typical projector that is lower bounded by 1 - $\delta$ by W.L.L.N giving $\sum^{n^l}_{j = 1}C_{\text{Max}}p(\psi^n_j)|\bra{\psi^n_j}\Lambda^n_\epsilon\ket{\psi^n_j}|^2 \geq C_{\text{Max}}(1 - 2\delta)$. The atypical compression contribution is formed by squared terms closer to 0 and so must be lower bounded differently, in particular a Cauchy-Schwarz type argument gives $\sum^{n^l}_{j = 1}C_{\text{Max}}p(\psi^n_j)|\bra{\psi^n_j}{\Lambda^n_\epsilon}^\perp \ket{\psi^n_j}|^2 \geq (1~-~C_{\text{Max}})\delta^2$. The contribution from the guess state can be neglected by positivity and additivity of the fidelity giving the bound presented above $\blacksquare$. In Fig.~\ref{fig:plot} this fidelity bound is labeled as $\overline{\mathcal{F}}_\text{Probe}$. 

In Appendix~\ref{appendix:proof}, we derive this inequality in full giving a tighter version for a fixed choice of guess state. We may equivalently model the impact of a thermal measurement probe as a noisy channel $\mathcal{N}$ with a single Kraus operator $K = (\sqrt{C_{\text{Max}}} - \sqrt{1 -C_{\text{Max}}})\mathbbm{1}_n$ applied to the instrument $\mathcal{T}^{\epsilon}_n(\cdot)$  (see Appendix~\ref{appendix:noisy}). Using $\mathcal{N}$ we find an expression for the impact of a thermal measurement probe over the set of $n$ sample messages drawn from the Haar measure 
\begin{align}
    \overline{\mathcal{F}}(\mathcal{N}) = \frac{2^{2n}\left(\sqrt{C_{\mathrm{Max}}^{(\beta, H_A,d)}} - \sqrt{1 - C_{\mathrm{Max}}^{(\beta, H_A,d)}}\right)^2 + 2^n}{2^{2n} + 2^n}. \label{eq:alice_fidelity}
\end{align} Treating, the imperfect measurement as a noisy channel we also obtain a modified gentle operator lemma~\cite{winter,wilde_2013}.
\begin{corollary}
(Roughly Gentle Lemma)
 Given a pure state message $\ket{\psi^n}$ formed of $n$ qubit samples, a typical projector $\Lambda^\epsilon_n$ and a noisy channel on the measurement $\mathcal{N}(\cdot)$ we have
\begin{align}
 \left|\left|\ketbra{\psi^n}{\psi^n} - \mathcal{N}(\mathcal{T}^\epsilon_n(\ketbra{\psi^n}{\psi^n}))\right|\right|_1 \leq 2\sqrt{\delta} + (1 - C_{\mathrm{Max}}^{(\beta, H_A,d)}). \nonumber
\end{align}
\end{corollary}
This follows from an application of the triangle inequality on the trace norm and the gentle operator lemma~\cite{winter}. The proof in detail is given in Appendix~\ref{appendix:noisy}. This corollary can be used as a means to prove an imperfect coding theorem in the style of Winter~\cite{winter} and highlights the impact of a thermal probe, where even if the message is large and $\delta\rightarrow 0$, $C_{\text{Max}}^{(\beta, H_A,d)}$ still contributes showing the impact of statistical mixing.

\section{Imperfectly timed encoding \& decoding.} 
\label{sec:timing}
Encoding and decoding in Schumacher compression involves rotating the message space using $U_f$ such that the eigenstates spanning the typical subspace are concentrated in the first $\lceil n S(\rho_\chi) + \epsilon\rceil$ qubits whilst the remainder are discarded. The compressed state $\ket{\psi^n}_c$ is then communicated to Bob who appends J qubits in the $\ket{0}$ state to $\ket{\psi^n}_c$ which is decoded using $U^\dagger_f$. Since $U_f$ and $U^\dagger_f$ are inverse unitaries, they are both generated by applying the same Hamiltonian $H_f$ for a fixed time $\tau$. If both processes generating the encoding and decoding are timed by identical and independent clocks with a Gaussian tick distribution~\cite{schwarzhans,meier2023} with mean $\tau$ and variance $\sigma$ as in~\cite{xuereb2023}, one will have concatenated dephasing in the energy eigenbasis of $H_f$. Examining such an evolution term by term in the energy eigenbasis of $H_f$ one notes that the time evolution of $U_f$ is still negated by $U^\dagger_f$ despite the impact of dephasing, which now occurs twice. As a result, imperfectly timed encoding and decoding results in an exponential envelope impacting the fidelity of the ensemble state Bob reconstructs $\mathcal{F}(\ket{\psi^n},\widetilde{\rho_R}) = \mathcal{F}_\text{Clocks}$ where
\begin{align}
\mathcal{F}_\text{Clocks} = \sum_{j,k} &e^{-\sigma^2(E_k - E_j)^2}\nonumber\\
 &\braket{k|\psi^n}\hspace{-4pt}\braket{\psi^n|j}\hspace{-4pt}\bra{\psi^n}\Lambda^\epsilon_n\ket{k}\hspace{-4pt}\bra{j}\Lambda^\epsilon_n\ket{\psi^n}. \label{eq:timing}
\end{align}
This means that the impact is only multiplicative depending on the tick variance $\sigma$. On the other hand, imperfect typical measurement can be much more harmful to the compression protocol since it not only degrades the fidelity but also introduces an additive error as atypical message states are at times erroneously
passed on when they should not be. The interested reader may find a full derivation for Eq.\eqref{eq:timing} in Appendix~\ref{appendix:timing}.
\section{Decoding with thermal qubits.}
\label{sec:decoding}
When decoding, Bob appends $n - \lceil n S(\rho_X) + \epsilon\rceil) = J$ qubits in the $\ket{0}$ state to the state they receive, which in the perfect measurement case is $\ket{\psi^n_c}$. They then decode by applying $U^\dagger_f$ to $\ket{\psi^n_c}\otimes \ket{0}^{ \otimes J}$ recovering $\ket{\psi^n}$. But the third law of thermodynamics points to the fact that pure states require diverging resources to produce~\cite{Taranto_2023,Buffoni_2022}. As such for a realistic thermodynamic analysis of quantum coding, let us allow Bob access to qubit thermal states of the Hamiltonian $H = - \hbar\omega \sigma_z$ of the form $\rho_{\ket{0}}(\eta) = (1 - \eta)\ketbra{0}{0} + \eta\ketbra{1}{1}$ which have a ground state population $1-\eta$ close to unity, but which one could produce using finite resources. To understand the impact of appending thermal states as opposed to pure states in decoding we compare the decompressed pure state $\ket{\psi^n}$ obtained by Bob with access to $\ket{0}^{\otimes J}$ with $\nu = U^\dagger_f\left(\ketbra{\psi^n_c}{\psi^n_c} \otimes \rho_{\ket{0}}(\eta)^{\otimes J}\right) U_f$ using the fidelity $\mathcal{F}(\ket{\psi^n},\nu) = \mathcal{F}_\text{append}$
\begin{align}
    \mathcal{F}_\text{append} = \bra{\psi^n}\nu\ket{\psi^n} = (1-\eta)^J, 
\end{align}
a full derivation is provided in Appendix~\ref{appendix:decoding}. Since $1 - \eta$ is a population of a thermal state of the Hamiltonian $H = -\hbar\omega \sigma_z$ i.e. $1-\eta = 1/2(1 + \tanh (\beta \hbar\omega))$, for a fixed energy gap $\omega$ we can rewrite this fidelity as an expression of the inverse temperature $\beta$ of the qubit states Bob has access to, the length of the message $n$ and the entropy of the source as 
\begin{align}
    \mathcal{F}_\text{append} &= \frac{1}{2^J}\left(1 + \tanh(\beta \hbar\omega)\right)^{n - \lceil n S(\rho_X) + \epsilon\rceil)}, \label{eq:bob_fid}
\end{align}
which exposes the fact that under realistic constraints the more Alice is able to compress, the harder it will be to achieve a good decoding fidelity, since Bob has to append more qubits and their non-zero temperature will decrease the fidelity exponentially in the number of appended qubits $J$.
\section{Entropy Produced by Alice \& Bob.}
\label{sec:entropy_production}
For a qubit measurement probe, the correlations that Alice is able to form with the probe to successfully carry out typical measurement are directly related to the ground state population of the qubit they have access to $C_{\text{Max}}^{(\beta, H_A,d)}~=~1 - \eta,$ and it is these correlations which constrain the achievable fidelity as we have seen in prior sections. Similarly, as we can see from Eq.\eqref{eq:bob_fid} the fidelity Bob can achieve when decoding is directly limited by the ground state populations of the qubits they have access to. But, \textit{what if we allow our agents to make use of the thermal states they have access to obtain more resourceful states?} A strategy they could take is to generate colder states with higher ground state populations and so achieve better fidelities. This is possible through a quantum thermodynamic cooling protocol~\cite{Silva_2016,Clivaz_2019_bound} where populations are exchanged between a reservoir of thermal qubits and a target qubit via a scheme by Reeb and Wolf~\cite{Reeb_2014}. A more detailed introduction to cooling for state preparation in this context is presented in Appendix~\ref{appendix:cooling}. It follows from Theorem 2 of~\cite{meier2023energyconsumption} that for the scheme proposed in~\cite{Reeb_2014}, an agent with access to $\tau_\beta$ and a reservoir $R = \bigotimes^{L}_{i=1}\tau_{\beta,\omega_i}$ with $L$ qubits in the state $\tau_\beta$ but with increasing energy gap $\omega_i$ can cool their system down to $\rho_{\ket{0}} (\eta)$ by swapping populations between $\tau_\beta$ and a constituent of the reservoir in 
\begin{align}L = \left\lceil \frac{(e+1)}{e \kappa} \ln \left(\frac{(e+1)}{\eta \kappa}\right) \right\rceil =\mathcal{O}\left(\frac{1}{\kappa}\log \frac{4}{\kappa \eta}\right) \label{eq:meier}
\end{align}
steps with an entropy production of at most $\kappa$, i.e. $\langle \Sigma \rangle = \beta \Delta Q + \Delta S \leq \kappa$. We can invert Eq.\eqref{eq:meier} to gain an insight into the minimum entropy production required to obtain the desired $\rho_{\ket{0}}(\eta)$ in the prescribed $L$ steps.
\begin{corollary}
\label{cor_1}
Given access to a reservoir $R = \bigotimes^L_{i = 1} \tau_{\beta, \omega_i}$ of $L$ qubits in a thermal state $\tau_\beta$ an agent can cool a target qubit initially in the state $\tau_\beta$ to $\rho_{\ket{0}}(\eta)$ in $L$ steps interacting collisionally with $R$, through a protocol having an entropy production $\langle \Sigma \rangle$ that satisfies
\begin{align}
  1 -  \frac{1}{\langle \Sigma \rangle} e^{-L \langle \Sigma \rangle} \geq 1 - \eta. \label{eq:ent_bound}
\end{align}
\end{corollary}
To see that this holds, from Eq.\eqref{eq:meier} we obtain $\eta = \frac{(e + 1)}{\kappa}e^{-(e/e+1)L\kappa}$ and since we have the inequalities $e^{-x} \leq e^{-a x}$ for $x \geq 0$ and $0~\leq~a~\leq~1$ as well as $\langle \Sigma \rangle \leq \kappa$ the above holds as a corollary of Theorem 2 of~\cite{meier2023energyconsumption}.

This readily implies that Bob's achievable fidelity for decoding Eq.\eqref{eq:bob_fid} is upper bounded by a function of the entropy produced to cool the qubits they append to the compressed state
\begin{align}
    \mathcal{F}(\ket{\psi^n},\nu) = (1-\eta)^J \leq  \left(1 -  \frac{1}{\langle \Sigma \rangle} e^{-L \langle \Sigma \rangle}\right)^J.
\end{align}
The Haar-averaged achievable fidelity in Alice's imperfect typical measurement Eq.\eqref{eq:alice_fidelity} that is $\overline{\mathcal{F}}(\mathcal{N}) = \frac{2^{2n}\left(\sqrt{1 - \eta} - \sqrt{\eta}\right)^2 + 2^n}{2^{2n} + 2^n}$ for the case of a qubit probe where $C_{\text{Max}}^{(\beta, H_A,d)}~=~1 - \eta$,  may be similarly upper bounded. Here we find
\begin{align}
    \overline{\mathcal{F}}(\mathcal{N}) \leq \frac{2^{2n}g(L,\langle \Sigma \rangle)^2 + 2^n}{2^{2n} + 2^n}
\end{align}
where this average fidelity is upper bounded in terms of a function of the entropy produced to cool the qubit probe Alice uses and the number of steps in the protocol to cool the probe, where $g(L,\langle \Sigma \rangle)= \sqrt{1 -  \frac{1}{\langle \Sigma \rangle} e^{-L \langle \Sigma \rangle}} -\sqrt{\frac{1}{\langle \Sigma \rangle} e^{-L \langle \Sigma \rangle}}.$ A detailed account of these derivations is given in Appendix~\ref{appendix:cooling}.

Using a result from Wilming \& Gallego's equality form of the 3rd Law of quantum thermodynamics~\cite{wilming_gallego} applied to our setting
\begin{gather}
T^{(L)}_S = \frac{1}{L}\frac{\langle H \rangle_\beta}{D(\rho_\beta || \tau_{\beta_R})},
\end{gather} we can also examine how the number of qubits Bob requires $L$ scales in an optimal cooling scenario with the desired fidelity. Where in the above expression the qubit state $\rho_{\ket{0}}(\eta)$ with the desired ground state population $1 - \eta$ has been expressed as a thermal state with a desired temperature $T^{(L)}_S = 1/\beta$. In Appendix~\ref{appendix:no_qubits} we find that the no. of qubits Bob requires to cool down the qubits they wish to append scales with the desired fidelity as 
\begin{align}
    L = \mathcal{O}\left( \ln \left(\frac{\mathcal{F}^{1/J}}{1 - \mathcal{F}^{1/J}}\right)\right).
\end{align}
Whilst this scaling seems logarithmic at first glance, the function is of the form $\ln(1+x/1-x)$ which corresponds to an inverse hyperbolic tangent. This means that the number of qubits $L$ Bob needs access to to cool down the thermal qubits $J$ they have access to and will use to append and decode the message scales in a hyperbolic way with the desired fidelity $\mathcal{F}_\text{append}$. This being said, different cooling protocols have different scaling relationships. In particular cooling protocols that make use of cooperative effects have been known to scale better than those that make use of collisional schemes~\cite{rolandi_23,Buffoni2023cooperativequantum} that we've made use of~\cite{Reeb_2014,meier2023energyconsumption,wilming_gallego}. This points to the fact that quantum thermodynamics plays a key role in investigating the scaling of resources in quantum protocols and that its consideration in considering the complexity of a quantum information protocol is a promising direction for future investigations.
\section{Discussion.}
\begin{figure}[t!]
    \centering
    \includegraphics[width = \columnwidth]{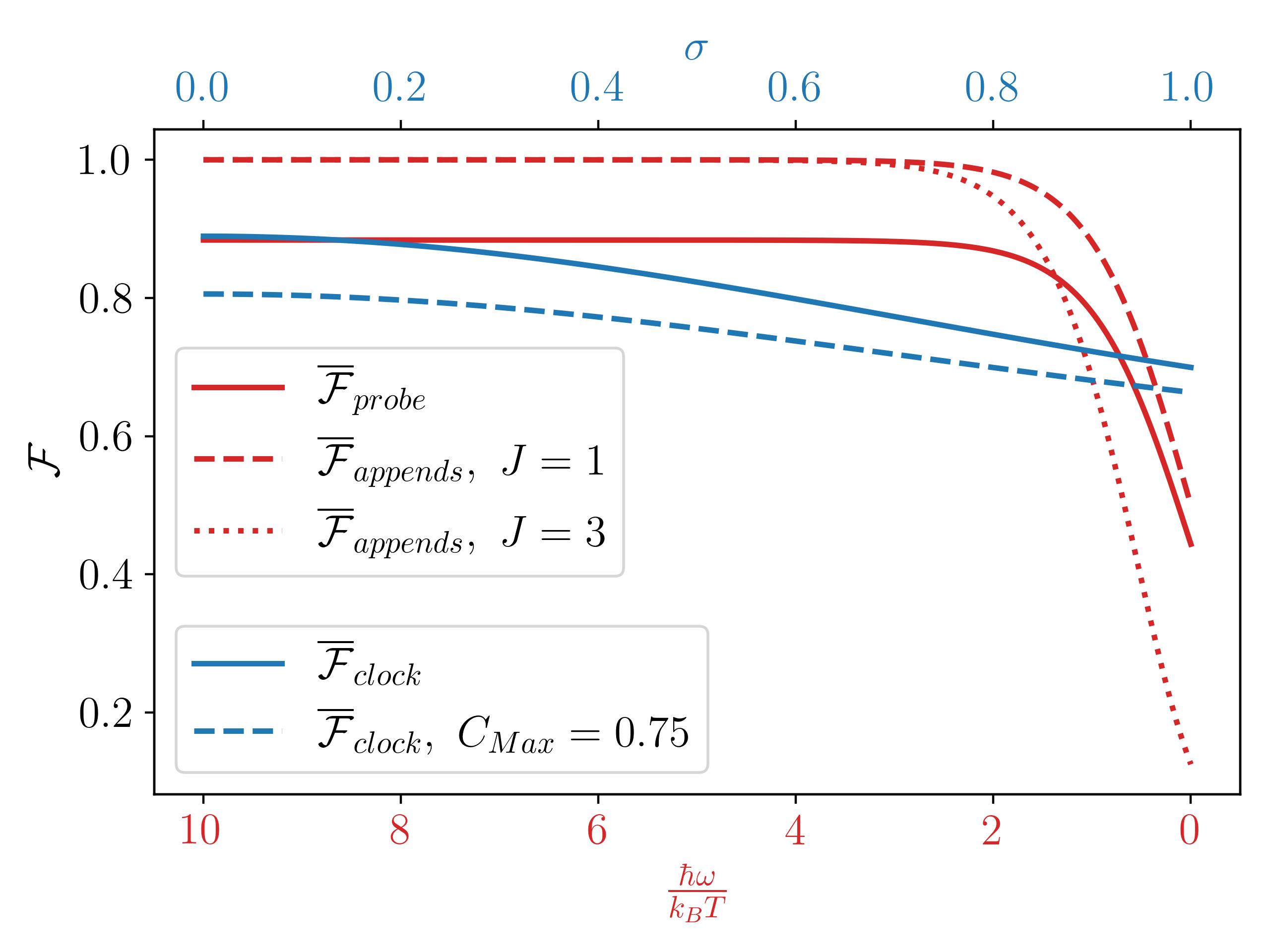}
    \caption{A plot of how the derived fidelity relationships scale against each other for a 3 qubit example as in Appenidx~\ref{appendix:instructive} with $\delta = 0.058$ and $\braket{\psi_G|\psi^3} \approx 0.79$, meaning that the perfect protocol fidelity for this example is around 0.92. In blue, we see the impact of imperfect timekeeping as the variance $\sigma$ of the tick distribution of the clocks which our agents make use of exponentially decreases the achievable fidelity (which cannot reach 1 as we exclude the guess state here). In red, we see the hyperbolic impact of a non-zero temperature in Alice's measurement probe and Bob's appended state which also exponentially decrease the fidelity in $J$ the number of appended states. $k_B$ is Boltzmann's constant and $T$ is temperature giving a sense for the energy scales considered. This plot can be generated using~\cite{code}.}
    \label{fig:plot}
\end{figure}

In the abstract of \cite{landauer}, Rolf Landauer remarks ``\textit{Information is not an abstract
entity but exists only through a physical representation, thus tying it to all the restrictions and
possibilities of our real physical universe.}''. In this work, we gave an account of how this insight holds within one of the foundational quantum information protocols, quantum coding. By showing that without access to diverging thermodynamic resources perfect quantum coding as presented by Schumacher~\cite{schumacher} is not achievable. While the scaling relationships we have derived show that large finite resource will allow an agent to get asymptotically close to perfect coding due to their hyperbolic form.
We achieved this by deriving fidelity relationships for various parts of the protocol with quantities which can be bounded by thermodynamic factors like the tick variance or ground state population. 

In particular, we modelled typical measurement as a unitary involving a thermal measurement probe and examined how this introduces statistical mixing which impacts the fidelity of the protocol unless the correlations formed with probe are perfect, something which requires diverging thermodynamic resources. Following this we examined how imperfect clocks doubly dephase encoding and decoding in the energy eigenbasis, introducing an exponential envelope to the achievable fidelity.  The last resource we examined was the ground state population or the temperature of the qubits appended by Bob, which introduce noise to the decoding process which is exponential in the number of appended qubits. This is different from the other thermodynamic impacts we examined as it scales with the goal of the protocol. I.e. the more Alice is able to compress, the more the protocol is susceptible to being impacted by error from the larger quantity of thermal qubits Bob will append.

In the last section of this work, we allow our agents to use the resources they have access to, to optimise them for this task through a quantum thermodynamic cooling procedure. Here we show that it is possible for them to improve the ground state population of the qubits they will use for measurement and appending, at the expense of entropy produced in the cooling protocol and thermal qubits to swap populations with. This cooling procedure unified the fidelity relationships we achieved earlier in the text, upper bounding them by the entropy produced in cooling to achieve a given fidelity. Thereby giving an operational interpretation to the fidelity expressions obtained and an insight into what the diverging thermodynamic resources represent.  

When examining how these different impacts scale against one another in Fig.~\ref{fig:plot} we see that the impact of the thermal measurement probe and appending a thermal qubit is the same, both are Hyperbolic in the temperature (although the probe fidelity is lower due to the statistical mixing introduced by atypical contributions). But once Alice starts to be able to compress more and so Bob is required to append more qubits, we see that it is the thermality of the appending qubits which impacts the fidelity the most as it scales exponentially in the number of qubits Alice discards. Finally, we saw that imperfect timekeeping adds a multiplicative envelope which is exponential in the variance of the ticks produced by Alice and Bob's clocks. It should be noted that is not all too bad for modern timekeeping devices which suffer from a $\sigma \approx 0.1-1$ ns due to electronic jitter~\cite{artiq_sinara,xuereb2023}.

In thinking about the compression of i.i.d.~quantum information thermodynamically, one might be tempted to draw comparisons with the deletion of quantum information. In Landauer erasure, heat is dissipated when information is destroyed, compensating for a change in entropy. In quantum coding, quantum information is communicated using the least number of qubits possible, begging the question of whether encoding information in a smaller Hilbert space results in heat dissipation. Quantum coding as introduced by Schumacher~\cite{schumacher} is completely unitary as one rotates the typical subspace such that it is concentrated over a number of qubits in the computational basis, meaning that the compression of quantum information itself does not introduce dissipation. It is the removal and deletion of the atypical subspace which results in dissipation due to Landauer erasure which was examined in~\cite{Vlatko}. The other rich thermodynamics we have examined, lies in the resources required to carry out the various parts of quantum compression, namely preparing states for measurement, encoding--decoding and timing processes to generate unitaries. 

\textit{Acknowledgements.---} 
The authors thank Max Lock, Faraj (Pharnam) Bakhshinezad and Toni Ac\'in for fruitful discussion. Florian Meier for help with tightening inequalities and Nicolai Friis for typesetting feedback. J.X. acknowledges that he has not been running enough and the support of Kooperativa Kixott. J.X., T.D., M.H. and P.E. would like to acknowledge funding from the European Research Council (Consolidator grant `Cocoquest’ 101043705). 
T.D. acknowledges support from {\"O}AW-JESH-Programme and the Brazilian agencies CNPq (Grant No. 441774/2023-7 and 200013/2024-6) and INCT-IQ through the project (465469/2014-0).
M.H. and P.E. further acknowledge funding by FQXi (FQXi- IAF19-03-S2, within the project “Fueling quantum field machines with information”) as well as the European flagship on quantum technologies (`ASPECTS' consortium 101080167). The views and opinions expressed are however those of the author(s) only
and do not necessarily reflect those of the European Union. Neither the European Union nor the granting authority can be held responsible for them. 

\bibliography{ref}% Produces the bibliography via BibTeX.

%\clearpage
\onecolumngrid
\appendix
\section*{Appendices}
\section{Three qubit example and some theory.}
\label{appendix:instructive}
Let us consider the textbook example \cite{Preskill1998,wilde_2013} of a quantum information source $\chi$ which outputs with probability $1/2$ the qubit state $\ket{0}$ or $\ket{+}$ at each interrogation. Alice samples 3 qubits from this source building a message which they wish to communicate to Bob using the least possible qubits. 

To begin with, the density of the source can be diagonalised to give $\rho_\chi = \cos^2 (\pi/8) \ketbra{0'}{0'} + \sin^2(\pi/8)\ketbra{1'}{1'}$ where $\ket{0'} = \cos(\pi/8) \ket{0} + \sin(\pi/8) \ket{1}$ and $\ket{1'} = \sin(\pi/8) \ket{0} + \sin(\pi/8)\ket{1}$. Let us use the notation $\eta_0 = \cos^2 (\pi/8)$ and $\gamma_1 = \sin^2(\pi/8)$ for brevity. Note that $\gamma_0 \approx 0.853$ and $\gamma_1 = 1 - \gamma_0$. Now a 3 sample message is obtained from the density $\rho^{\otimes 3}_\chi$ which has 8 eigenvalues of the form $\lambda_i = \binom{3}{i}\gamma_0^{3 -i}\gamma^{i}_1$, $i \in [0,3]$ where degeneracies have been summed over, meaning that the eigenstates contribute more to the density depending on the number of $\ket{0'}$ states which form them such that the $\ket{0'0'0'}$ state contributes most\footnote{One can think of this as a \textit{Hamming weight} and develop a notion of strong quantum typicality involving types \cite{wilde_2013} but we will not make use of this here. The states within a given type correspond to the degenerate eigenvalues and have identical labels up to permutation i.e. the states $\ket{0'0'1'} , \ket{0'1'0'}$ and $\ket{1'0'0'}$ share the eigenvalue $ \lambda_1 = 3 \gamma^2_0\gamma_1$.}. 

More generally, the eigenstates of $\rho^{\otimes n}_\chi$ generated by an i.i.d.~qubit source are binomially distributed according to their eigenvalues, so in the sense of Shannon compression when sampling from this binomial distribution one can obtain \textit{typical} and \textit{atypical} sequences of eigenvalues where by W.L.L.N. $\exists \, N \in \mathbb{N} \,:\, \forall n \geq N$ and $\epsilon, \delta > 0$ \begin{align}
     \mathbb{P}\left[\left|-\frac{1}{n}\sum^{2^n}_{i=1} \log \lambda_i - S(\rho_\chi)\right| > \epsilon \right] < \delta, \label{eq:WLLN}
\end{align} and we may call states with eigenvalues satisfying $|-\frac{1}{n}\sum^{2^n}_{i=1} \log \lambda_i - S(\rho_\chi)| \leq \epsilon$, $\epsilon$-typical. In this way, the projective Hilbert space of $n$ qubits may be partitioned into two subspaces corresponding to the states whose eigenvalues form $\epsilon$-typical sequences and those that do not. In this spirit we introduce the $\epsilon$-typical projector $\Lambda^\epsilon_n =\sum_{\lambda^n_k \in T^\epsilon_n} \ketbra{\psi^n_k}{\psi^n_k}$ comprised of the projectors of states $\ket{\psi^n_k}$ with eigenvalues $\lambda^n_k$ that are in the $\epsilon$-typical set of eigenvalues $T^\epsilon_n$. By virtue of eq.\eqref{eq:WLLN} and the definition of a typical projector we have that $\text{tr}\{\Lambda^\epsilon_n \rho^{\otimes n}_{\chi}\} \geq 1 - \delta$.

In the context of our example, for a suitable choice of $\epsilon$ we can can partition the three qubit Hilbert space into a subspace spanned by the eigenstates $\left\{\ket{0'0'0'},\ket{0'0'1'},\ket{0'1'0'},\ket{1'0'0'}\right\}$ defining the typical projector
\begin{align}
    \Lambda^{\epsilon}_3 = \ketbra{0'0'0'}{0'0'0'}+\ketbra{0'0'1'}{0'0'1'}+\ketbra{0'1'0'}{0'1'0'}+\ketbra{1'0'0'}{1'0'0'}
\end{align}
and naturally the atypical projector $\mathbbm{1} - \Lambda^{\epsilon}_3$ whose corresponding subspace is spanned by the remaining eigenstates. This allows us to extend the notion of $\epsilon$-typicality to states which lie prominently along the typical subspace. Alice, having sampled three qubits from from $\rho_\chi$ has obtained a pure state $\ketbra{\psi^3}{\psi^3}$ involving some combination of $\ket{0}$ and $\ket{+}$. Knowing the consequences of information theory, Alice is aware that if their message lies along the typical subspace of the source they only need $n(S(\rho_\chi)+\epsilon)$ qubits to encode their message as only roughly $2^{n(S(\rho_\chi)+\epsilon)}$ eigenvalues contribute to the information content. In this case, $n = 3$ and $S(\rho_\chi) \approx 0.6$ so Alice would need two qubits. Their strategy is now to carry out a typical measurement described by the channel \begin{gather}\mathcal{T}^{\epsilon}_3(\sigma) = \Lambda^\epsilon_3\sigma\Lambda^\epsilon_3 + (\mathbbm{1} - \Lambda^\epsilon_3)\sigma (\mathbbm{1} - \Lambda^\epsilon_3),
\end{gather}
if the $\ket{\psi^{3}}$ is successfully projected to the typical subspace then Alice rotates the basis of this state such that the relevant eigenstates are found on the Hilbert space of the first two qubits in the tensor product, allowing them to remove the third and communicate their message in two qubits to Bob. In this example, this rotation can be carried out by 
\begin{align}
  U_{\text{encode}} = \scalemath{0.6}{\left(
\begin{array}{cccccccc}
 \cos ^3\left(\frac{\pi }{8}\right) & \sin \left(\frac{\pi }{8}\right) \cos ^2\left(\frac{\pi }{8}\right) & \sin \left(\frac{\pi }{8}\right) \cos ^2\left(\frac{\pi }{8}\right) & \sin ^2\left(\frac{\pi }{8}\right) \cos \left(\frac{\pi }{8}\right) & \sin \left(\frac{\pi }{8}\right) \cos ^2\left(\frac{\pi }{8}\right) & \sin ^2\left(\frac{\pi }{8}\right) \cos \left(\frac{\pi }{8}\right) & \sin ^2\left(\frac{\pi }{8}\right) \cos \left(\frac{\pi }{8}\right) & \sin ^3\left(\frac{\pi }{8}\right) \\
 \sin \left(\frac{\pi }{8}\right) \cos ^2\left(\frac{\pi }{8}\right) & \sin ^2\left(\frac{\pi }{8}\right) \cos \left(\frac{\pi }{8}\right) & \sin ^2\left(\frac{\pi }{8}\right) \cos \left(\frac{\pi }{8}\right) & \sin ^3\left(\frac{\pi }{8}\right) & -\cos ^3\left(\frac{\pi }{8}\right) & \sin \left(\frac{\pi }{8}\right) \left(-\cos ^2\left(\frac{\pi }{8}\right)\right) & \sin \left(\frac{\pi }{8}\right) \left(-\cos ^2\left(\frac{\pi }{8}\right)\right) & \sin ^2\left(\frac{\pi }{8}\right) \left(-\cos \left(\frac{\pi }{8}\right)\right) \\
 \sin \left(\frac{\pi }{8}\right) \cos ^2\left(\frac{\pi }{8}\right) & \sin ^2\left(\frac{\pi }{8}\right) \cos \left(\frac{\pi }{8}\right) & -\cos ^3\left(\frac{\pi }{8}\right) & \sin \left(\frac{\pi }{8}\right) \left(-\cos ^2\left(\frac{\pi }{8}\right)\right) & \sin ^2\left(\frac{\pi }{8}\right) \cos \left(\frac{\pi }{8}\right) & \sin ^3\left(\frac{\pi }{8}\right) & \sin \left(\frac{\pi }{8}\right) \left(-\cos ^2\left(\frac{\pi }{8}\right)\right) & \sin ^2\left(\frac{\pi }{8}\right) \left(-\cos \left(\frac{\pi }{8}\right)\right) \\
 \sin ^2\left(\frac{\pi }{8}\right) \cos \left(\frac{\pi }{8}\right) & \sin ^3\left(\frac{\pi }{8}\right) & \sin \left(\frac{\pi }{8}\right) \left(-\cos ^2\left(\frac{\pi }{8}\right)\right) & \sin ^2\left(\frac{\pi }{8}\right) \left(-\cos \left(\frac{\pi }{8}\right)\right) & \sin \left(\frac{\pi }{8}\right) \left(-\cos ^2\left(\frac{\pi }{8}\right)\right) & \sin ^2\left(\frac{\pi }{8}\right) \left(-\cos \left(\frac{\pi }{8}\right)\right) & \cos ^3\left(\frac{\pi }{8}\right) & \sin \left(\frac{\pi }{8}\right) \cos ^2\left(\frac{\pi }{8}\right) \\
 \sin \left(\frac{\pi }{8}\right) \cos ^2\left(\frac{\pi }{8}\right) & -\cos ^3\left(\frac{\pi }{8}\right) & \sin ^2\left(\frac{\pi }{8}\right) \cos \left(\frac{\pi }{8}\right) & \sin \left(\frac{\pi }{8}\right) \left(-\cos ^2\left(\frac{\pi }{8}\right)\right) & \sin ^2\left(\frac{\pi }{8}\right) \cos \left(\frac{\pi }{8}\right) & \sin \left(\frac{\pi }{8}\right) \left(-\cos ^2\left(\frac{\pi }{8}\right)\right) & \sin ^3\left(\frac{\pi }{8}\right) & \sin ^2\left(\frac{\pi }{8}\right) \left(-\cos \left(\frac{\pi }{8}\right)\right) \\
 \sin ^2\left(\frac{\pi }{8}\right) \cos \left(\frac{\pi }{8}\right) & \sin \left(\frac{\pi }{8}\right) \left(-\cos ^2\left(\frac{\pi }{8}\right)\right) & \sin ^3\left(\frac{\pi }{8}\right) & \sin ^2\left(\frac{\pi }{8}\right) \left(-\cos \left(\frac{\pi }{8}\right)\right) & \sin \left(\frac{\pi }{8}\right) \left(-\cos ^2\left(\frac{\pi }{8}\right)\right) & \cos ^3\left(\frac{\pi }{8}\right) & \sin ^2\left(\frac{\pi }{8}\right) \left(-\cos \left(\frac{\pi }{8}\right)\right) & \sin \left(\frac{\pi }{8}\right) \cos ^2\left(\frac{\pi }{8}\right) \\
 \sin ^2\left(\frac{\pi }{8}\right) \cos \left(\frac{\pi }{8}\right) & \sin \left(\frac{\pi }{8}\right) \left(-\cos ^2\left(\frac{\pi }{8}\right)\right) & \sin \left(\frac{\pi }{8}\right) \left(-\cos ^2\left(\frac{\pi }{8}\right)\right) & \cos ^3\left(\frac{\pi }{8}\right) & \sin ^3\left(\frac{\pi }{8}\right) & \sin ^2\left(\frac{\pi }{8}\right) \left(-\cos \left(\frac{\pi }{8}\right)\right) & \sin ^2\left(\frac{\pi }{8}\right) \left(-\cos \left(\frac{\pi }{8}\right)\right) & \sin \left(\frac{\pi }{8}\right) \cos ^2\left(\frac{\pi }{8}\right) \\
 \sin ^3\left(\frac{\pi }{8}\right) & \sin ^2\left(\frac{\pi }{8}\right) \left(-\cos \left(\frac{\pi }{8}\right)\right) & \sin ^2\left(\frac{\pi }{8}\right) \left(-\cos \left(\frac{\pi }{8}\right)\right) & \sin \left(\frac{\pi }{8}\right) \cos ^2\left(\frac{\pi }{8}\right) & \sin ^2\left(\frac{\pi }{8}\right) \left(-\cos \left(\frac{\pi }{8}\right)\right) & \sin \left(\frac{\pi }{8}\right) \cos ^2\left(\frac{\pi }{8}\right) & \sin \left(\frac{\pi }{8}\right) \cos ^2\left(\frac{\pi }{8}\right) & -\cos ^3\left(\frac{\pi }{8}\right) \\
\end{array}
\right)}
\end{align}
which sends the eigenstates $\{\ket{0'0'0'},\ket{0'0'1'},\ket{0'1'0'},\ket{1'0'0'},\ket{0'1'1'},\ket{1'0'1'},\ket{1'1'0'},\ket{1'1'1'}\}$ to computational basis $\{\ket{000},\ket{110},\ket{010},\ket{100},\ket{101},\ket{011},\ket{001},\ket{111}\}$.

Bob would then append a $\ket{0}$ to the received message and decode the message using the inverse unitary. In the case that the message is measured to be atypical then it cannot be compressed and a guess state $\ket{\psi_G}$ is instead passed on to Bob. In total Bob would receive the ensemble 
\begin{gather}
    \rho_R = \Lambda^\epsilon_n\ketbra{\psi^n}{\psi^n}\Lambda^\epsilon_n + \bra{\psi^n}\left(\mathbbm{1} - \Lambda^\epsilon_n \right)\ket{\psi^n} \ketbra{\psi_G}{\psi_G},
\end{gather}
meaning that they obtain the message with a fidelity
\begin{align}
    \mathcal{F}\left(\ket{\psi^n}, \rho_R\right) = |\bra{\psi^n}\Lambda^\epsilon_n&\ket{\psi^n}|^2  + \bra{\psi}\left(\mathbbm{1} - \Lambda^\epsilon_n \right)\ket{\psi^n}|\braket{\psi_G|\psi^n}|^2.
\end{align}
The \textit{Mathematica} code associated to this example is available at~\cite{code}. 
\section{Typical Measurement with a Thermal Probe}
\subsection{von Neumann picture of measurement} In this setting, the quantum system of interest interacts unitarily with an ancillary system which is then projectively measured, giving statistics equivalent to measuring the system of interest. 

Consider a set of projection operators $\{\Pi_i\}$ such that the probability of observing outcome $x_i$ for a quantum state $\rho$ is $p(x_i) = \text{tr}\left\{\Pi_i \rho \right\}$ by the Born Rule. The von Neumann-L\"uders projection postulate \cite{petruccione} gives the state after measuring the outcome $x_i$ as 
\begin{gather}
    \rho_{x_i} = \frac{\Pi_i \rho \Pi_i}{\text{tr}\left\{\Pi_i \rho \right\}}
\end{gather}
meaning that the complete measurement induced by all the set of projectors gives a superposition of such states 
\begin{gather}
\rho'=\sum_i p(x_i) \rho_{x_i} = \sum_i \Pi_i \rho \Pi_i .
\end{gather} 
In the von Neumann picture of measurement, the probabilities describing the distribution of eigenstates are encoded into an ancillary system with dimension equal to the number of projectors of the measurement such that 
\begin{gather}
    \rho'_{AS} = \sum_i p(x_i) \ketbra{a_i}{a_i} \otimes \rho(x_i)
\end{gather}
and so projectively measuring the ancilla returns the measurement statistics of the system of interest indirectly as
\begin{gather}
\rho' = \sum_i p(x_i)\rho_{x_i}  = \text{tr}_A\left\{\rho'_{AS}\right\}.
\end{gather}
Mathematically, this picture can be seen as an application of the Naimark extension which relates a positive operator valued measurement on the Hilbert space of some system to a unitary interaction with an ancillary system in a dilated Hilbert Space followed by a projective measurement of this ancillary system.
\subsection{Typical measurement in the von Neumann picture}
The typical measurement is binary establishing whether the spectrum of a quantum state representing a message is a typical sequence or not, and so involves two projectors $\Pi_i = \left\{\Lambda^\epsilon_n, \mathbbm{1} -\Lambda^\epsilon_n\right\}$. As such we only need to add two additional degrees of freedom to our message Hilbert Space of size $2^n$ for $n$ samples, in the form of an ancilla qubit $\rho_A \, \in \, \mathcal{H}_A$. The ensemble post measurement state $\rho'_M$ can thus be expressed via this dilation as 
\begin{align}
\rho'_M &= \text{tr}_A\left\{V\rho_M \otimes \rho_A V^\dagger\right\} \label{eq:dilation} \\
&= \Lambda^\epsilon_n \rho_M \Lambda^\epsilon_n + \left(\mathbbm{1} -\Lambda^\epsilon_n\right) \rho_M \left(\mathbbm{1} -\Lambda^\epsilon_n\right) \label{eq:kraus} \\
&= \rho'_{M|0} + \rho'_{M|1},
\end{align}
where we've projected onto the eigenbasis of the ancilla to recover the measurement statistics of the message. I have also introduced the notation $\rho'_{M|i}$ to indicate the conditional post-measurement state of the system given that the ancilla is found in state $i$. The object of interest to us is the unitary interaction within this dilation. For the sake of inspecting further we stipulate that the ancilla is in the state $\rho_A = \ketbra{0}{0}$ in eq.~\ref{eq:dilation}
\begin{gather}
\rho'_M = \bra{0} V \ket{0} \rho_M \bra{0} V^\dagger \ket{0} + \bra{1} V \ket{0} \rho_M \bra{0} V^\dagger \ket{1}
\end{gather}
allowing us to identify the first column of $V$ in block form
\begin{gather}
    V = \begin{pmatrix} \Lambda^\epsilon_n & \rvline &  A\\
\hline
  \mathbbm{1} -\Lambda^\epsilon_n & \rvline & B
\end{pmatrix}
\end{gather}
whilst the second column is unknown with blocks $A$ and $B$ which we assume to be Hermitian. To satisfy unitarity we require that $VV^\dagger = V^\dagger V = \mathbbm{1}$ which gives the equations 
\begin{align}
&\Lambda^\epsilon_n + A^2 = \mathbbm{1}\\
&B^2 - \Lambda^\epsilon_n = \mymathbb{0}
\end{align}
that constrain the form of $A$ and $B$. One such solution is
\begin{align}
    V &= \begin{pmatrix} \Lambda^\epsilon_n & \rvline &  \mathbbm{1} - \Lambda^\epsilon_n\\
\hline
  \mathbbm{1} - \Lambda^\epsilon_n & \rvline &  \Lambda^\epsilon_n 
\end{pmatrix},
 \label{eq:unitary}
\end{align}
where in short we have $V = \Lambda^\epsilon_n \otimes \mathbbm{1}_A + \left(\mathbbm{1} - \Lambda^\epsilon_n\right) \otimes \sigma_x$ as the dilated unitary on the message state and probe. In this way, if the the probe is in the ground state its ground state population after the unitary is related to the probability of the message being in the typical subspace.

\subsection{Weakening the typical measurement} Consider the typical measurement in the von Neumann picture as describe before. Our goal in this section is to see how robust compression is against faulty typical measurement by considering imperfect von Neumann measurements where the imperfection is a result of the mixedness of the ancilla, which we will impose to be thermal.

The desired averaged post-measurement state of a typical measurement is 
\begin{gather}
\rho'_M = \Lambda^\epsilon_n \rho_M \Lambda^\epsilon_n + \left(\mathbbm{1} -\Lambda^\epsilon_n\right) \rho_M \left(\mathbbm{1} -\Lambda^\epsilon_n\right)    
\end{gather}
and as discussed in the previous section, this can obtained via a unitary process $V = \Lambda^\epsilon_n \otimes I_1 + (\mathbbm{1}-\Lambda^\epsilon_n)\otimes \sigma^x_1$ on the code state and a probe ancilla placed in the ground state. Preparing this ancilla in the ground state is a contentious matter that has been used to attribute infinite cost associated to carrying out perfect measurements. To see how much the typical measurement is affected by the state of the probe let us instead consider this ancilla to be in a thermal state $\tau_\beta = \frac{e^{-\beta H}}{\mathcal{Z}}$ of the qubit Hamiltonian $H = -\hbar\omega \sigma_z$ at inverse  temperature $\beta$. The ancilla now gives a more general ensemble post-measurement state
\begin{align}
    \rho'_M &= \text{tr}_A\left\{V \rho_M \otimes \tau_\beta V^\dagger\right\} \\
            &= \frac{e^{-\beta E_1}}{\mathcal{Z}}\left(\bra{0}V\ket{0}\rho_M\bra{0}V^\dagger\ket{0} + \bra{1}V\ket{0}\rho_M\bra{0}V^\dagger\ket{1}\right) \\
            &+ \frac{e^{-\beta E_2}}{\mathcal{Z}}\left(\bra{0}V\ket{1}\rho_M\bra{1}V^\dagger\ket{0} + \bra{1}V\ket{1}\rho_M\bra{1}V^\dagger\ket{1}\right) \\
            &= \frac{e^{-\beta E_1}}{\mathcal{Z}}\left( \Lambda^\epsilon_n \rho_M \Lambda^\epsilon_n + (\mathbbm{1}-\Lambda^\epsilon_n ) \rho_M ( \mathbbm{1}-\Lambda^\epsilon_n ) \right) + \frac{e^{-\beta E_2}}{\mathcal{Z}}\left( ( \mathbbm{1}-\Lambda^\epsilon_n ) \rho_M ( \mathbbm{1}-\Lambda^\epsilon_n )  + \Lambda^\epsilon_n \rho_M \Lambda^\epsilon_n \right)
\end{align}
where we note that the ensemble post-measurement state is identical but that the conditional measurement states are different such that we have 
\begin{align}
    \rho'_{M|0} &= \frac{e^{-\beta E_1}}{\mathcal{Z}}\Lambda^\epsilon_n\rho_M{\Lambda^\epsilon_n} + \frac{e^{-\beta E_2}}{\mathcal{Z}}(\mathbbm{1}-\Lambda^\epsilon_n)\rho_M(\mathbbm{1}-\Lambda^\epsilon_n) \nonumber \\
    \rho'_{M|1}&= \frac{e^{-\beta E_1}}{\mathcal{Z}}(\mathbbm{1}-\Lambda^\epsilon_n)\rho_M(\mathbbm{1}-\Lambda^\epsilon_n)+ \frac{e^{-\beta E_2}}{\mathcal{Z}}\Lambda^\epsilon_n\rho_M{\Lambda^\epsilon_n}  \label{eq:new_kraus}
\intertext{as opposed to}
    \rho'_{M|0} &= \Lambda^\epsilon_n\rho_M{\Lambda^\epsilon_n}\\
    \rho'_{M|1} &=(\mathbbm{1}-\Lambda^\epsilon_n)\rho_M(\mathbbm{1}-\Lambda^\epsilon_n).
\end{align}
The typical measurement statistics are recovered either when the qubit probe is cooled to the ground state i.e. $\beta \to \infty$ or it has an infinite energy gap i.e. $E_1 \to \infty$. When the qubit measurement probe does not fulfil either of these criteria, the probe has an \textit{a priori} bias which impacts the measurement statistics and as a result impacts our ability to compress, as we are able to identify less of the message with its projection into the typical subspace which is then encoded. Note that this observation echoes the sentiment of~\cite{Guryanova_2020} where it is shown that perfect projective measurements require diverging resources -- here we have shown that as a result of the insights of \cite{Guryanova_2020} Schumacher compression requires diverging resources for perfect typical measurement.
\paragraph{Strategies to mitigate non-zero temperature}
A strategy taken in \cite{Guryanova_2020} to mitigate the dependence on temperature and the spectrum of the Hamiltonian of the thermal probe is to introduce another diverging quantitiy, probe size. Consider now enlarging the probe to a $d$ dimensional qudit thermal state with Hamiltonian $H_A = \sum^d_{i=1} E_i \ketbra{i}{i}$. We apply the same unitary structure which implements the typical measurement in the von Neumann picture but now extended via the unitary
\begin{gather}
    V = \Lambda^\epsilon_n \otimes \mathbbm{1}_d + (\mathbbm{1}_n - \Lambda^\epsilon_n) \otimes X_{d/2}
\end{gather}
where $X_{d/2} = \sum^{d/2 - 1}_{j = 0} \ketbra{E_{j + d/2}}{E_j} + \ketbra{E_j}{E_{j + d/2}}$ which conditioned on the state of the qubit carries an effective \textit{NOT} gate on the $d$-level thermal state, swapping its first $d/2$ populations with its last $d/2$ energy eigenstates.
This results in the conditional post-measurement state
\begin{gather}
\rho'_{M|0} = \sum^{d/2 - 1}_{j = 0}\frac{e^{-\beta E_j}}{\mathcal{Z}}\Lambda^\epsilon_n\rho_M{\Lambda^\epsilon_n} + \sum^{d}_{k = d/2}\frac{e^{-\beta E_k}}{\mathcal{Z}}(\mathbbm{1}-\Lambda^\epsilon_n)\rho_M(\mathbbm{1}-\Lambda^\epsilon_n) \label{eq:cond_pm_state}
\end{gather}
where the sum of the first half of the populations of the qudit thermal state correspond to the probability of the message state being typical and the sum of the second half correspond to the probability of the message state being atypical. In the limit of $d \to \infty$ the typical measurement statistics are again recovered since $\sum^{d/2 - 1}_{j = 0}\frac{e^{-\beta E_j}}{\mathcal{Z}} \to 1$ and $\sum^{d}_{k = d/2}\frac{e^{-\beta E_k}}{\mathcal{Z}} \to 0$ as $d \to \infty$.

\subsection{Average Fidelity of Compression with Imperfect Typical Measurement} \label{appendix:proof}
Alice samples $n$ times from a source $\chi$ with $\rho_\chi = \sum^l_{i = 1} p_i \ketbra{\phi_i}{\phi_i}$ per sample obtaining the message $\ket{\psi^n}$ which she compresses and sends to Bob after imperfect typical measurement with a probe that forms correlations $C_{\text{Max}}^{(\beta, H_A,d)}$ with the state. Bob receives the statistical mixture
\begin{align}
    \widetilde{\rho_R} = C_{\text{Max}}^{(\beta, H_A,d)} \Lambda^\epsilon_n \ketbra{\psi^n}{\psi^n}\Lambda^\epsilon_n  &+ \left(1 - C_{\text{Max}}^{(\beta, H_A,d)}\right)\left(\mathbbm{1} - \Lambda^\epsilon_n\right) \ketbra{\psi^n}{\psi^n}(\mathbbm{1} - \Lambda^\epsilon_n) \nonumber \\
   &+ \left(\left(1 - C_{\text{Max}}^{(\beta, H_A,d)}\right)\bra{\psi^n}\Lambda^\epsilon_n\ket{\psi^n} + C_{\text{Max}}^{(\beta, H_A,d)}\bra{\psi^n}\left(\mathbbm{1} - \Lambda^\epsilon_n\right)\ket{\psi^n}\right)\ketbra{\psi_G}{\psi_G}
\end{align}
where $\ket{\psi_G}$ is a chosen guess state. The fidelity of this mixed state with the pure message state $\ket{\psi^n}$ is
\begin{align}
    \mathcal{F}(\widetilde{\rho_R},\ket{\psi^n}) &= \bra{\psi^n}\widetilde{\rho_R}\ket{\psi^n} \\
    &= C_{\text{Max}}^{(\beta, H_A,d)} |\bra{\psi^n}\Lambda^\epsilon_n \ket{\psi^n}|^2 + \left(1 - C_{\text{Max}}^{(\beta, H_A,d)}\right)|\bra{\psi^n}\left(\mathbbm{1} - \Lambda^\epsilon_n\right)\ket{\psi^n}|^2 \nonumber \\ &+ \left(\left(1 - C_{\text{Max}}^{(\beta, H_A,d)}\right)\bra{\psi^n}\Lambda^\epsilon_n\ket{\psi^n} + C_{\text{Max}}^{(\beta, H_A,d)}\bra{\psi^n}\left(\mathbbm{1} - \Lambda^\epsilon_n\right)\ket{\psi^n}\right)|\braket{\psi_G|\psi}|^2.
\end{align}
Consider the set of possible $n$ sample messages $\{\ket{\psi^n_j}\}^{n^l}_{j=1}$ each occurring with probability $p(\psi^n_j)$ then the fidelity of the received state with the message state averaged over the set of possible message is 
\begin{align}
    \overline{\mathcal{F}} &= \sum^{n^l}_{j = 1} p(\psi^n_j) \left( 
    C_{\text{Max}}^{(\beta, H_A,d)} |\bra{\psi^n_j}\Lambda^\epsilon_n \ket{\psi^n_j}|^2 + \left(1 - C_{\text{Max}}^{(\beta, H_A,d)}\right)|\bra{\psi^n_j}\left(\mathbbm{1} - \Lambda^\epsilon_n\right)\ket{\psi^n_j}|^2 \right. \nonumber \\ 
    &+ \left. \left(\left(1 - C_{\text{Max}}^{(\beta, H_A,d)}\right)\bra{\psi^n_j}\Lambda^\epsilon_n\ket{\psi^n_j} + C_{\text{Max}}^{(\beta, H_A,d)}\bra{\psi^n_j}\left(\mathbbm{1} - \Lambda^\epsilon_n\right)\ket{\psi^n_j}\right)|\braket{\psi_G|\psi^n_j}|^2 \right).
\end{align}
Noting that $\text{tr}\{\Lambda^\epsilon_n \rho^{\otimes n}\} = \sum^{n^l}_{j = 1} p(\psi^n_j)\bra{\psi^n_j}\Lambda^\epsilon_n\ket{\psi^n_j} \geq 1 - \delta$ and similarly $\text{tr}\{\left(\mathbbm{1} - \Lambda^\epsilon_n\right)\rho^{\otimes n}\} \geq \delta$ we have that the guess state term is lower bounded by
\begin{align}
\overline{\mathcal{F}} &\geq \sum^{n^l}_{j = 1} p(\psi^n_j) \left( 
    C_{\text{Max}}^{(\beta, H_A,d)} |\bra{\psi^n_j}\Lambda^\epsilon_n \ket{\psi^n_j}|^2 + \left(1 - C_{\text{Max}}^{(\beta, H_A,d)}\right)|\bra{\psi^n_j}\left(\mathbbm{1} - \Lambda^\epsilon_n\right)\ket{\psi^n_j}|^2 \right) \nonumber \\
    &+  \left(\left(1 - C_{\text{Max}}^{(\beta, H_A,d)}\right)(1 - \delta) + C_{\text{Max}}^{(\beta, H_A,d)}\delta\right)|\braket{\psi_G|\psi^n_j}|^2.
\end{align}
We may begin to lower bound this average fidelity by noting that since $|\bra{\psi^n_j}\Lambda^\epsilon_n \ket{\psi^n_j}|^2$ is close to 1 for small $\delta$, this ensures that the lower bound $x^2 \geq 2x - 1$ is not loose for $x = \bra{\psi^n_j}\Lambda^\epsilon_n \ket{\psi^n_j}$ giving 
\begin{align}
\overline{\mathcal{F}} &\geq \sum^{n^l}_{j = 1} p(\psi^n_j) \left( 
    C_{\text{Max}}^{(\beta, H_A,d)}\left(2\bra{\psi^n_j}\Lambda^\epsilon_n \ket{\psi^n_j}| - 1\right) + \left(1 - C_{\text{Max}}^{(\beta, H_A,d)}\right)|\bra{\psi^n_j}\left(\mathbbm{1} - \Lambda^\epsilon_n\right)\ket{\psi^n_j}|^2 \right) \nonumber \\
    &+  \left(\left(1 - C_{\text{Max}}^{(\beta, H_A,d)}\right)(1 - \delta) + C_{\text{Max}}^{(\beta, H_A,d)}\delta\right)|\braket{\psi_G|\psi^n_j}|^2 \\
    &=C_{\text{Max}}^{(\beta, H_A,d)}\left(1 - 2\delta\right) + \left(1 - C_{\text{Max}}^{(\beta, H_A,d)}\right)\sum^{n^l}_{j = 1} p(\psi^n_j)|\bra{\psi^n_j}\left(\mathbbm{1} - \Lambda^\epsilon_n\right)\ket{\psi^n_j}|^2 \nonumber \\ 
    &+ \left(\left(1 - C_{\text{Max}}^{(\beta, H_A,d)}\right)(1 - \delta) + C_{\text{Max}}^{(\beta, H_A,d)}\delta\right)|\braket{\psi_G|\psi^n_j}|^2. \label{eq:b34}
\end{align}
Fixing our focus on the second and most problematic term, it cannot be simplified like the first term since it is close to~0. Instead consider that $\mathbbm{R}^{n^l}$ may be equipped with an inner product $\langle \, , \, \rangle_{M}$ such that $\forall \, \mathbf{v}, \mathbf{w} \, \in \mathbbm{R}^{n^l}, \, \langle \mathbf{v}, \mathbf{w}\rangle_{M} = \mathbf{v}^{T}M\mathbf{w}$ for some positive, semi-definite matrix $M \in \mathbbm{R}^{n^l \times n^l}$. This allows us to think of $\delta$ as 
\begin{align}
    \delta  = \sum_{j = 1} p(\psi^n_j) \bra{\psi^n_j}\left(\mathbbm{1} - \Lambda^\epsilon_n\right)\ket{\psi^n_j} = \langle \mathbf{f} , \mathbf{1}\rangle_{P}
\end{align}
where $\mathbf{f} = \sum^{n^l}_{j = 1} \bra{\psi^n_j}\left(\mathbbm{1} - \Lambda^\epsilon_n\right)\ket{\psi^n_j} \mathbf{e}_j$, $\mathbf{1} = \sum^{n^l}_{j = 1} \mathbf{e}_j$ are vectors formed by $\left\{\mathbf{e}_j\right\}^{n^l}_{j = 1}$ basis vectors in $\mathbbm{R}^{n^l}$. And $P = \sum^{n^l}_{j} \mathbf{e_j}\mathbf{e_j}^{T} p(\psi^n_j)$ is a diagonal matrix in $\mathbbm{R}^{n^l \times n^l}$. The Cauchy - Schwarz inequality may then be employed 
\begin{align}
\delta = \langle \mathbf{f} , \mathbf{1}\rangle_{P} &\leq \sqrt{\langle \mathbf{f},\mathbf{f}\rangle_P \langle \mathbf{1},\mathbf{1}\rangle_P} \\
&= \sqrt{\left(\sum^{n^l}_{j = 1} p(\psi^n_j)|\bra{\psi^n_j}\left(\mathbbm{1} - \Lambda^\epsilon_n\right)\ket{\psi^n_j}|^2 \right)\left(\sum^{n^l}_{k = 1} p(\psi^n_k)\right)} \\
&= \sqrt{\sum^{n^l}_{j = 1} p(\psi^n_j)|\bra{\psi^n_j}\left(\mathbbm{1} - \Lambda^\epsilon_n\right)\ket{\psi^n_j}|^2}
\end{align}
which implies the following lower bound on the term of interest  
\begin{align}
\delta^2 \leq \sum^{n^l}_{j = 1} p(\psi^n_j)|\bra{\psi^n_j}\left(\mathbbm{1} - \Lambda^\epsilon_n\right)\ket{\psi^n_j}|^2
\end{align}
and we may write \eqref{eq:b34} as 
\begin{align}
    \overline{\mathcal{F}} \geq 
    C_{\text{Max}}^{(\beta, H_A,d)}\left(1 - 2\delta\right) + \left(1 - C_{\text{Max}}^{(\beta, H_A,d)}\right)\delta^2 + \left(\left(1 - C_{\text{Max}}^{(\beta, H_A,d)}\right)(1 - \delta) + C_{\text{Max}}^{(\beta, H_A,d)}\delta\right)\sum^{n^l}_{j = 1} p(\psi^n_j)|\braket{\psi_G|\psi^n_j}|^2.
\end{align}
By positivity we may neglect the guess state term giving the inequality presented in the main text
\begin{align}
    \overline{\mathcal{F}} \geq 
    C_{\text{Max}}^{(\beta, H_A,d)}\left(1 - 2\delta\right) + \left(1 - C_{\text{Max}}^{(\beta, H_A,d)}\right)\delta^2,
\end{align}which reduces to the conventional bound $\overline{\mathcal{F}} \geq 1 - 2\delta$ for $C_{\text{Max}}^{(\beta, H_A,d)} = 1$. $\blacksquare$
\subsection{Representing Imperfect Measurement as a Noisy Channel} \label{appendix:noisy} To examine the impact of a thermal probe on measurement we can reformulate the above calculation as a noisy channel concatenated with our desired instrument eq.\eqref{eq:typ_meas} allowing us to bound the average fidelity of the resultant imperfect measurement with the ideal case. Consider that $\mathcal{T}^{\epsilon}_n(\cdot)$ has Kraus operators $K_1 = \Lambda^\epsilon_n$ and $K_2 = \mathbbm{1} - \Lambda^\epsilon_n$ which by comparing with eq.\eqref{eq:new_kraus} we see are transformed as 
\begin{align}
    \widetilde{K_1} \longrightarrow \sqrt{\alpha}K_1 + i\sqrt{1 - \alpha}K_2 && \widetilde{K_2} \longrightarrow \sqrt{\alpha}K_2 + i\sqrt{1-\alpha}K_1 
\end{align}
where $\alpha = \sum^{d/2 - 1}_{j = 0}\frac{e^{-\beta E_j}}{\mathcal{Z}} =C_{\text{Max}}^{(\beta, H_A,d)}$. This transformation can be obtained by concatenating $\mathcal{T}^{\epsilon}_n(\cdot)$ with the noisy channel $\mathcal{N}$ with a single Kraus operator $K = (\sqrt{\alpha} + \sqrt{1 - \alpha})\mathbbm{1}$, which whilst being unitary and preserving the ensemble sum of populations impacts the post-measurement outcomes as in eq.\eqref{eq:cond_pm_state}. To see that this channel is trace preserving consider
\begin{gather}
    \widetilde{K}^\dagger_1\widetilde{K}_1 = \alpha K_1^\dagger K_1 + i\sqrt{1-\alpha}\sqrt{\alpha} K^\dagger_2 K_1 -i \sqrt{1-\alpha}\sqrt{\alpha} K_1^\dagger K_2 + (1 - \alpha)K^\dagger_2 K_2 = \alpha \Lambda^\epsilon_n + (1 - \alpha)(\mathbb{1} - \Lambda^\epsilon_n) \\
    \widetilde{K}^\dagger_2\widetilde{K}_2 = \alpha K_2^\dagger K_2 - i\sqrt{1-\alpha}\sqrt{\alpha} K^\dagger_1 K_2 + i\sqrt{1-\alpha}\sqrt{\alpha} K_2^\dagger K_1 + (1 - \alpha)K^\dagger_1 K_1 = \alpha (\mathbb{1} - \Lambda^\epsilon_n) +  (1 - \alpha)\Lambda^\epsilon_n
\end{gather}
where we have used the fact that $K_1 = \Lambda^\epsilon_n$ and $K_2 = \mathbb{1} - \Lambda^\epsilon_n$ are Hermitian and idempotent since they are projectors and that $K_1$ and $K_2$ commute. This gives the completeness relation $\widetilde{K}^\dagger_1\widetilde{K}_1 + \widetilde{K}^\dagger_2\widetilde{K}_2 = \mathbb{1}$ as desired.

It was shown in~\cite{Nielsen_2002,Emerson_2005} that the average fidelity is dependent only on the noisy channel $\mathcal{E}(\cdot) = \sum_i M_i (\cdot)M_i^\dagger$ being considered as $
\overline{\mathcal{F}}(\mathcal{E}) = \frac{\sum_i |\text{tr} K_i|^2 + d}{d^2 +d},$
where $d$ is the dimension of the Hilbert space we are averaging over. This means that the average fidelity of an imperfectly measured state with a perfectly desired state may be expressed as 
\begin{align}
    \overline{\mathcal{F}}(\mathcal{N}) = \frac{2^{2n}\left(\sqrt{C_{\text{Max}}^{(\beta, H_A,d)}} - \sqrt{1 - C_{\text{Max}}^{(\beta, H_A,d)}}\right)^2 + 2^n}{2^{2n} + 2^n}, 
\end{align}
where we have Haar-averaged over all possible $n$ qubit messages.
\begin{corollary}
(Roughly Gentle Lemma)
 Given a pure state message $\ket{\psi^n}$ formed of $n$ qubit samples, a typical projector $\Lambda^\epsilon_n$ and a noisy channel on the measurement $\mathcal{N}(\cdot)$ we obtain a modified gentle operator lemma~\cite{winter,wilde_2013} as
 \begin{align}
 \left|\left|\ketbra{\psi^n}{\psi^n} - \mathcal{N}(\mathcal{T}^\epsilon_n(\ketbra{\psi^n}{\psi^n}))\right|\right|_1 \leq 2\sqrt{\delta} + (1 - C_{\mathrm{Max}}^{(\beta, H_A,d)}). \nonumber
 \end{align}
\end{corollary}
To begin consider the LHS and add and subtract $\Lambda^\epsilon_n\ketbra{\psi^n}{\psi^n}\Lambda^\epsilon_n$
\begin{align}
    \left|\left|\ketbra{\psi^n}{\psi^n} - \mathcal{N}(\mathcal{T}^\epsilon_n(\ketbra{\psi^n}{\psi^n}))\right|\right|_1 &=  \left|\left|\ketbra{\psi^n}{\psi^n} - \Lambda^\epsilon_n\ketbra{\psi^n}{\psi^n}\Lambda^\epsilon_n + \Lambda^\epsilon_n\ketbra{\psi^n}{\psi^n}\Lambda^\epsilon_n - \mathcal{N}(\mathcal{T}^\epsilon_n(\ketbra{\psi^n}{\psi^n}))\right|\right|_1\\
\intertext{which allows us to make use to triangle inequality on the trace distance as}
    &\leq \left|\left|\ketbra{\psi^n}{\psi^n} - \Lambda^\epsilon_n\ketbra{\psi^n}{\psi^n}\Lambda^\epsilon_n\right|\right|_1 + \left|\left| \Lambda^\epsilon_n\ketbra{\psi^n}{\psi^n}\Lambda^\epsilon_n - \mathcal{N}(\mathcal{T}^\epsilon_n(\ketbra{\psi^n}{\psi^n}))\right|\right|_1\\
    &= \left|\left|\ketbra{\psi^n}{\psi^n} - \Lambda^\epsilon_n\ketbra{\psi^n}{\psi^n}\Lambda^\epsilon_n\right|\right|_1 \nonumber\\ 
    &+ \left|\left|\Lambda^\epsilon_n\ketbra{\psi^n}{\psi^n}\Lambda^\epsilon_n-C_{\text{Max}}\Lambda^\epsilon_n\ketbra{\psi^n}{\psi^n}\Lambda^\epsilon_n - (1-C_{\text{Max}})(\mathbbm{1}-\Lambda^\epsilon_n)\ketbra{\psi^n}{\psi^n}(\mathbbm{1}-\Lambda^\epsilon_n)\right|\right|_1 \\
    &\leq 2\sqrt{\delta} + (1 - C_{\text{Max}})\left|\left|\Lambda^\epsilon_n\ketbra{\psi^n}{\psi^n}\Lambda^\epsilon_n - (\mathbbm{1}-\Lambda^\epsilon_n)\ketbra{\psi^n}{\psi^n}(\mathbbm{1}-\Lambda^\epsilon_n)\right|\right|_1\\
    \intertext{where we have used the gentle operator lemma~\cite{winter,wilde_2013} and collected terms. The second term is the trace norm of a valid quantum state which is equal to unity allowing us to state the result }
    \left|\left|\ketbra{\psi^n}{\psi^n} - \mathcal{N}(\mathcal{T}^\epsilon_n(\ketbra{\psi^n}{\psi^n}))\right|\right|_1&\leq 2\sqrt{\delta} + (1-C_{\text{Max}}^{(\beta, H_A,d)}). \, \blacksquare
\end{align}

\section{Imperfect Encoding and Decoding with identical clocks dephases twice}
\label{appendix:timing}
In Schumacher Compression, Alice encodes their $n$ qubit message state $\ket{\psi^n}$ after verifying that it is typical via an encoding unitary $U_f$ as $U_f\ket{\psi^n} = \ket{\psi^n_c} \otimes\ket{\psi_{\text{junk}}}$. After this $\ket{\psi_{\text{junk}}}$ is traced out and $\ket{\psi^n_c}$ is communicated to Bob who appends $J$ qubits in the $\ket{0}$ state and decodes the compressed state $U^{\dagger} \ket{\psi^n_c}\otimes\ket{0}^{\otimes J} = \ket{\psi^n}$. Alice and Bob carry out these encoding and decoding operation by timing some physical process described by the same Hamiltonian since $U_f$ and $U_f^{\dagger}$ are generated by the same Hamiltonian say $H_f$, for the same time say $\tau$. 

Let Alice and Bob each have access to identical and independent clocks having Gaussian tick distributions with 1st moment $\tau$ and 2nd moment $\sigma$. If they are successful in coding and decoding, the concatenated action of $U_f$, $U_f^{\dagger}$ would be the identity. Instead we have the channel
\begin{align}
    \mathcal{D}_2(\rho) &= \int^{\infty}_{-\infty} \frac{e^{\frac{(t’-\tau)^2}{-2\sigma^2}}}{\sqrt{2\pi\sigma^2}} U_f^\dagger(t') \int^{\infty}_{-\infty} \frac{e^{\frac{(t-\tau)^2}{-2\sigma^2}}}{\sqrt{2\pi\sigma^2}} U_f(t) \rho U_f^\dagger(t)\, dt \, U_f(t') \, dt' \\
    &=\int^{\infty}_{-\infty} \frac{e^{\frac{(t’-\tau)^2}{-2\sigma^2}}}{\sqrt{2\pi\sigma^2}} e^{i H_f t'} \int^{\infty}_{-\infty} \frac{e^{\frac{(t-\tau)^2}{-2\sigma^2}}}{\sqrt{2\pi\sigma^2}} e^{-i H_f t} \rho e^{i H_f t}\, dt \, e^{- i H_f t'} \, dt',
\end{align}
let $H_f = \sum^{2^n}_{j=1} E_j\ketbra{j}{j}$ in the energy eigenbasis then term by term this expression becomes
\begin{align}
 \bra{j} \mathcal{D}_2(\rho) \ket{k} &= \int^{\infty}_{-\infty} \frac{e^{\frac{(t’-\tau)^2}{-2\sigma^2}}}{\sqrt{2\pi\sigma^2}} e^{ -i(E_k - E_j)t'} \int^{\infty}_{-\infty} \frac{e^{\frac{(t-\tau)^2}{-2\sigma^2}}}{\sqrt{2\pi\sigma^2}} e^{-i(E_j - E_k)t} \bra{j}\rho \ket{k} \, dt \,\, dt', \\
 &= e^{\frac{-\sigma^2}{2} (E_k - E_j)^2 } e^{-i(E_k - E_j)\tau}e^{\frac{-\sigma^2}{2}(E_j - E_k)^2}e^{-i(E_j - E_k)\tau} \bra{j} \rho \ket{k} \\
 &= e^{\frac{-\sigma^2}{2}(E_k - E_j)^2}e^{\frac{-\sigma^2}{2}(E_j - E_k)^2} \bra{j}\rho \ket{k} \\
 &= e^{-\sigma^2(E_k - E_j)^2}\bra{j}\rho \ket{k}
\end{align}
and so imperfect concatenated encoding and decoding leads to dephasing twice in the energy eigenbasis of the Hamiltonian that generates the encoding and decoding unitary. This gives the following fidelity for Bob's recovered ensemble state with Alice's sampled message state
\begin{align} \mathcal{F}(\ket{\psi^n},\widetilde{\rho_R}) &= \sum_{j,k} e^{-\sigma^2(E_j - E_k)^2} \braket{j|\psi}\braket{\psi|k} \left[C_{\text{Max}}\bra{\psi}\Lambda\ket{j}\bra{k}\Lambda\ket{\psi^n} + (1- C_{\text{Max}})\bra{\psi}\mathbbm{1} - \Lambda\ket{l}\bra{k}\mathbbm{1} - \Lambda\ket{\psi^n}\right] + \nonumber \\
 &\left[C_{\text{Max}}\bra{\psi}\mathbbm{1} - \Lambda\ket{\psi^n} + (1-C_{\text{Max}})\bra{\psi}\Lambda\ket{\psi^n}\right]\left|\braket{\psi|\psi_G}\right|^2.
\end{align}

\section{Appending Thermal States}
\label{appendix:decoding}
When decoding, Bob appends $n - \lceil n S(\rho_X) + \epsilon\rceil) = J$ qubits in the $\ket{0}$ state to the state they receive, which in the perfect measurement case is $\ket{\psi^n_c}$. They then decode by applying $U^\dagger_f$ to $\ket{\psi^n_c}\otimes \ket{0}^{ \otimes J}$ recovering $\ket{\psi^n}$. But the third law of thermodynamics points to the fact that pure states require diverging resources to produce~\cite{Taranto_2023,Buffoni_2022} as such for a realistic thermodynamic analysis of quantum coding, let us allow Bob access to qubit thermal states of the Hamiltonian $H = - \hbar\omega \sigma_z$ of the form $\rho_{\ket{0}}(\eta) = (1 - \eta)\ketbra{0}{0} + \eta\ketbra{1}{1}$ which are close to the ground state, but which one could produce using finite resources. To understand the impact of appending thermal states as opposed to pure states in decoding we compare the decompressed pure state $\ket{\psi^n}$ obtained by Bob with access to $\ket{0}^{\otimes J}$ with $\nu = U^\dagger_f \ketbra{\psi_c}{\psi_c} \otimes \rho_{\ket{0}}(\eta)^{\otimes J} U_f$ using the fidelity
\begin{align}
    \mathcal{F}(\ket{\psi^n},\nu) &= \bra{\psi^n}\nu\ket{\psi^n} \\
    &= \bra{\psi^n} U^\dagger_f \ketbra{\psi_c}{\psi_c} \otimes \rho_{\ket{0}}(\eta)^{\otimes J} U_f \ket{\psi^n} \\
    &= \bra{0}^{\otimes J}\otimes \bra{\psi_c} U_f U^\dagger_f \ketbra{\psi_c}{\psi_c} \otimes \rho_{\ket{0}}(\eta)^{\otimes J} U_f U^\dagger_f \ket{\psi^n_c}\otimes \ket{0}^{\otimes J}\\
\intertext{where we have used $\ket{\psi^n} = U^\dagger_f \left(\ket{\psi^n_c}\otimes \ket{0}^{\otimes J}\right)$ which by unitarity and distributivity of the inner product over the tensor product gives}
 &= \bra{0}^{\otimes J}\otimes \bra{\psi_c}\left(\ketbra{\psi_c}{\psi_c} \otimes \rho_{\ket{0}}(\eta)^{\otimes J}\right)\ket{\psi^n_c}\otimes \ket{0}^{\otimes J}\\
&= \bra{0}^{\otimes J}\rho_{\ket{0}}(\eta)^{\otimes J}\ket{0}^{\otimes J} |\braket{\psi_c|\psi_c}|^2 \\
&= \bra{0}^{\otimes J}\left((1 - \eta)\ketbra{0}{0} + \eta\ketbra{1}{1}\right)^{\otimes J}\ket{0}^{\otimes J}\\
\intertext{here, the only non-zero overlap will come from the all zero kets $\ket{0}^{\otimes J}$ by orthonormality giving}
\mathcal{F}(\ket{\psi^n},\nu) &= (1-\eta)^J. \label{eq:ap_bob_fidelity}
\end{align}
Since $1 - \eta$ is a population of a thermal state of the Hamiltonian $H = -\hbar\omega \sigma_z$ i.e. $1-\eta = 1/2(1 + \tanh (\hbar\beta \omega))$ we can rewrite this fidelity as an expression of the inverse temperature $\beta$ of the qubit states Bob has access to, the size of the message $n$ and the entropy of the source as 
\begin{align}
    \mathcal{F}(\ket{\psi^n},\nu) &= \frac{1}{2^J}\left(1 + \tanh(\hbar\beta \omega)\right)^{n - \lceil n S(\rho_X) + \epsilon\rceil)} \\
    &=\frac{1}{\mathcal{Z}^J}\left(1 - e^{-\hbar\beta E_1}\right)^{n - \lceil n S(\rho_X) + \epsilon\rceil)},
\end{align}
which exposes the fact that under realistic constraints the more you compress, the more you pay in terms of fidelity since Bob has to append more qubits and their impurity will start to contribute more and more. 

\section{Entropy Produced by Alice and Bob}
\begin{figure}
    \centering
\includegraphics[width = 0.6\textwidth]{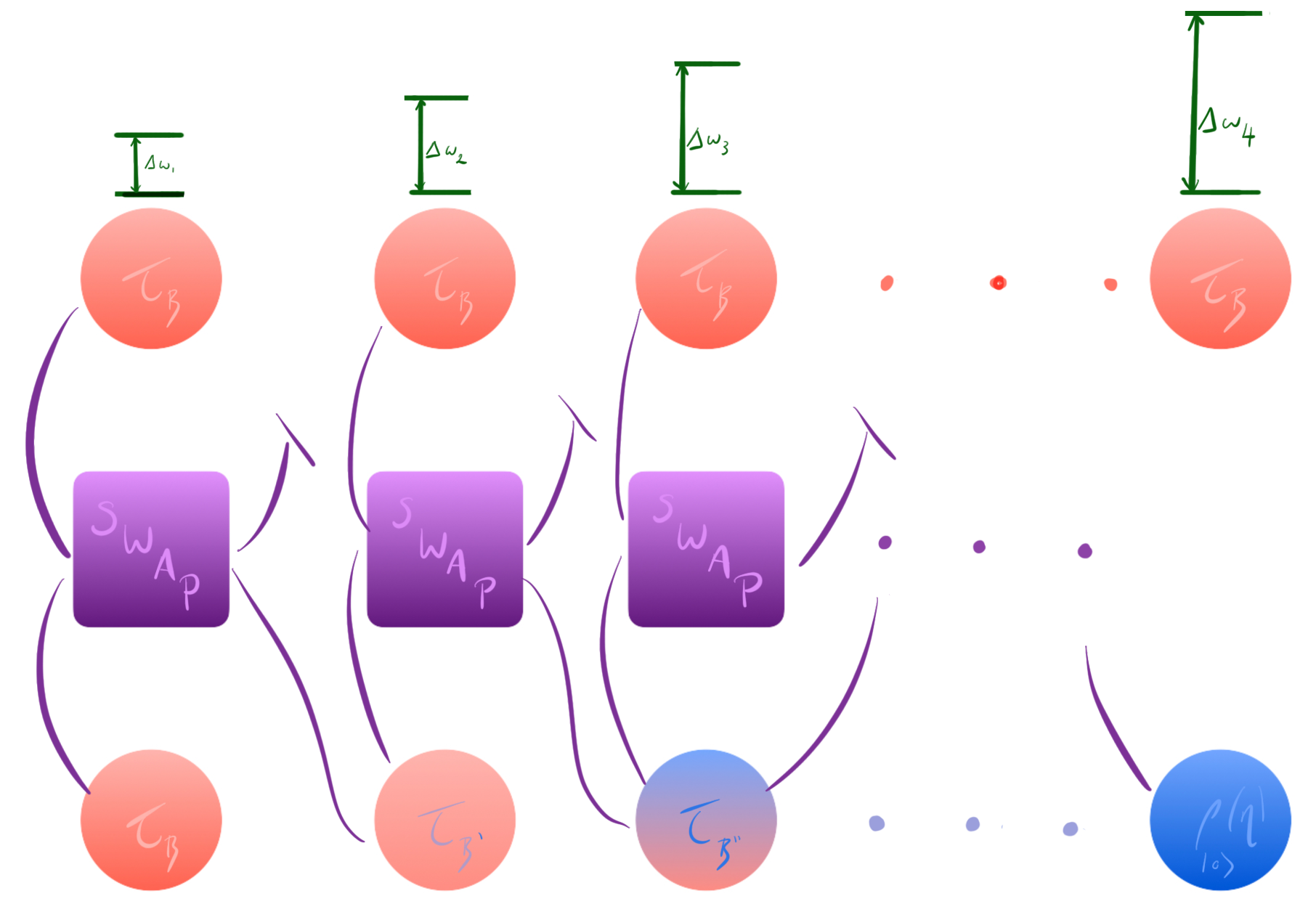}
    \caption{An illustration depicting the described cooling protocol.}\label{fig:cooling_illustration}
\end{figure}
\label{appendix:cooling}
In carrying out this compression protocol Alice and Bob both need access to some resourceful states, in particular they needs access to states as close to the $\ket{0}$ state as possible. For Alice, they require $d/2$ qubits close to the $\ket{0}$ state to construct the $d$ dimensional qudit probe they will use to determine whether the message of pure states they have sampled is typical or not. For Bob, they require $J$ qubits as close to the $\ket{0}$ state as possible so as to be able to append these to the state they receive $\ket{\psi^n_c}$ and decode this compressed state to recover the original message $\ket{\psi^n}$ with a high fidelity. 

To account for the resources required to generate these resourceful states, let us assume that to begin with Alice and Bob both have access to a large $D$ dimensional environment $\rho_E$ at temperature $\beta$ and qubits which are in a thermal state $\tau_\beta$ at temperature $\beta$ with this environment. Their goal would then be to carry out some global unitary protocol $G$ on the target qubit and environment such that at the end they obtain the desired state close to $\ket{0}$ as
\begin{align}
    \rho_{\ket{0}}(\eta) = \text{tr}\left\{G \tau_\beta \otimes \rho_E G^{\dagger} \right\}.
\end{align}
This setup is known as Landauer erasure or cooling~\cite{Reeb_2014,Silva_2016,Clivaz_2019_bound,Taranto_2023} within the field of quantum thermodynamics~\cite{thermo_review, binder_book}. In particular, a protocol to cool a target quantum system in this manner was introduced by Reeb and Wolf~\cite{Reeb_2014} where the environment $\rho_E$ is in a separable state formed of several identical thermal subsystems with the same dimension as the target system. An agent then swaps populations between the target and one of the subsystems of the environment at each step $t$ of a protocol which results in $V$. In our setting, let us assume that Alice and Bob have access to an environment $\rho_E =  \bigotimes^{L}_{i=1}\tau_{\beta,\omega_i}$ with $L$ qubits in the state $\tau_\beta$ but with increasing energy gap $\omega_i$. And they wish to cool their target qubit $\tau_\beta$ to $\rho_{\ket{0}}(\eta)$. Theorem 2 of~\cite{meier2023energyconsumption} by Meier and Yamasaki shows that this is possible using a finite $L$ dimensional environment in finite time steps $L$ with entropy production $\langle \Sigma \rangle$ which is upper bounded by a constant $\kappa$. 

Entropy production $\langle \Sigma \rangle$ is a quantity related to the heat a system dissipates into the environment $\beta\Delta Q$ and its change in entropy $\Delta S$ where $\langle \Sigma \rangle = \beta\Delta Q + \Delta S$. Another informative formulation of the entropy production~\cite{Esposito_2010,Reeb_2014} is in terms of the mutual information $\mathcal{I}(S':E')$ formed between a system and an environment and the relative entropy of the initial and final states of the environment $D(E'||E)$ 
\begin{align}
    \langle \Sigma \rangle = \mathcal{I}(S':E') + D(E'||E),
\end{align}
which applied to the context of cooling would give one a sense for how correlated the system being cooled and the environment become as well as how much the environment heats up during the protocol.

It follows from Theorem 2 of~\cite{meier2023energyconsumption} that for an agent with access to $\tau_\beta$ and an environment ${\tau_\beta}^{\otimes L}$ they can cool their system down to $\rho_{\ket{0}} (\eta)$ by swapping populations between $\tau_\beta$ and a constituent of the environment in \begin{align}L = \left\lceil \frac{(e+1)}{e \kappa} \ln \left(\frac{(e+1)}{\eta \kappa}\right) \right\rceil =\mathcal{O}\left(\frac{1}{\kappa}\log \frac{4}{\kappa \eta}\right) \label{eq:ap_meier} \end{align} steps with an entropy production of at most $\kappa$, i.e. $\langle \Sigma \rangle = \beta \Delta Q + \Delta S \leq \kappa$. This means that to cool the qubit thermal state an amount of entropy must be produced depending on how hot the environment is allowed to get and how large its dimension is (which is equal to the number of operations an agent would need for this cooling protocol). In this way, we are able to state that Alice produces and entropy production of at most $d\kappa$ to obtain the measurement probe and carry out the typical measurement probe and Bob expends $J\kappa$ entropy production to decode the compressed state.

We can invert eq.\eqref{eq:ap_meier} to make a statement which is more relevant for our setting which is a that a function of the entropy production lower bounds the achievable distance from unit ground state population. Giving us an idea for the minimum entropy production required to obtain $\rho_{\ket{0}}(\eta)$ in the prescribed $L$ steps. From eq.\eqref{eq:ap_meier}  we obtain 
\begin{align}
    \eta = \frac{(e + 1)}{\kappa}e^{-(\frac{e}{e+1})L\kappa}
\end{align}
and since we have the inequalities $e^{-x} \leq e^{-a x}$ for $x \geq 0$ and $0 \leq a \leq 1$ and the bounded entropy production $\langle \Sigma \rangle \leq \kappa$ we may state as a corollary of Theorem 2 of~\cite{meier2023energyconsumption} that to obtain a qubit ground state population of $1 - \eta$ in $L$ steps one requires an entropy production satisfying
\begin{align}
    \frac{1}{\langle \Sigma \rangle} e^{-L \langle \Sigma \rangle} \leq \eta, \label{eq:ap_ent_bound_2}
\end{align}
or alternatively the achievable ground state population is upper bounded 
\begin{align}
  1 -  \frac{1}{\langle \Sigma \rangle} e^{-L \langle \Sigma \rangle} \geq 1 - \eta. \label{eq:ap_ent_bound}
\end{align}
This readily implies that Bob's achievable fidelity for decoding eq.\eqref{eq:ap_bob_fidelity} is upper bounded by a function of the entropy produced to cool the qubits they append to the compressed state
\begin{align}
    \mathcal{F}(\ket{\psi^n},\nu) = (1-\eta)^J \leq  \left(1 -  \frac{1}{\langle \Sigma \rangle} e^{-L \langle \Sigma \rangle}\right)^J.
\end{align}
The achievable fidelity in Alice's imperfect typical measurement eq.\eqref{eq:alice_fidelity} may be similarly upper bounded by a function of the entropy production spent to cool the probe for the case of a qubit probe where $C_{\text{Max}}^{(\beta, H_A,d)} = 1 - \eta$ that is
\begin{align}
    \overline{\mathcal{F}}(\mathcal{N}) = \frac{2^{2n}\left(\sqrt{1 - \eta} - \sqrt{\eta}\right)^2 + 2^n}{2^{2n} + 2^n}.
\end{align}
Now consider that $\left(\sqrt{1 - \eta} - \sqrt{\eta}\right)^2 = \left(\sqrt{\eta} - \sqrt{1-\eta}\right)^2$ and taking the square root of both sides of eq.\eqref{eq:ap_ent_bound} 
\begin{gather}
  \sqrt{1 -  \frac{1}{\langle \Sigma \rangle} e^{-L \langle \Sigma \rangle}} \geq \sqrt{1 - \eta}
\end{gather}
from which we can subtract $\sqrt{\eta}$ from both sides
\begin{align}
  \sqrt{1 -  \frac{1}{\langle \Sigma \rangle} e^{-L \langle \Sigma \rangle}} -\sqrt{\eta}\geq \sqrt{1 - \eta} - \sqrt{\eta},\\
\intertext{and from eq.\eqref{eq:ap_ent_bound_2} we have that $\sqrt{\frac{1}{\langle \Sigma \rangle} e^{-L \langle \Sigma \rangle}} \leq \sqrt{\eta}$ then it is clear that}
\underbrace{\sqrt{1 -  \frac{1}{\langle \Sigma \rangle} e^{-L \langle \Sigma \rangle}} -\sqrt{\frac{1}{\langle \Sigma \rangle} e^{-L \langle \Sigma \rangle}}}_{= g(L,\langle \Sigma \rangle)}\geq \sqrt{1 - \eta} - \sqrt{\eta}.
\end{align}
Squaring both sides of this inequality we find an upper bound on Alice's Haar averaged fidelity of imperfect measurement over the set of possible messages 
\begin{align}
    \overline{\mathcal{F}}(\mathcal{N}) \leq \frac{2^{2n}g(L,\langle \Sigma \rangle)^2 + 2^n}{2^{2n} + 2^n}
\end{align}
in terms of a function of the entropy produced to cool the qubit probe they use and the number of step in the protocol to cool the probe.

\subsection{How many thermal qubits does Bob need to achieve a given fidelity?}
\label{appendix:no_qubits}
While Corollary~\ref{cor_1} quantifies the no. of qubits $L$ required to cool the target qubit to a ground state population $1 - \eta$ at a fixed entropy production, one could alternatively ask how the no. of thermal qubits $L$ at a temperature $\beta$ scales asymptotically with to achieve a given fidelity close to 1. Physically, this question boils down to the limits of cooling a quantum system and so the third law of quantum thermodynamics~\cite{wilming_gallego, Taranto_2023}. In~\cite{wilming_gallego} Wilming and Gallego propose a quantitative form of the third law of quantum thermodynamics described by an equality. In our setting, Bob needs to cool $J$ thermal qubits to append and replace the traced out junk qubits and we can use the insights of~\cite{wilming_gallego} to examine the scaling of the resources required in this scenario. Bob's cooling protocol corresponds to the I.I.D cooling case presented in \cite{wilming_gallego} where a resource state $\rho_R = (\tau_{\beta_R})^{\otimes L}$ composed of $L$ constituent identical systems is used to cool a system of interest of dimension equal to the constituents, to a temperature $T_S$. This result is derived in the context of the resource theory of thermodynamics~\cite{Lostaglio_2019} using energy preserving unitaries, where the system of interest and resource state are embedded in an environment at temperature $\beta_E$. In this context, the scaling relationship
\begin{gather}
    T^{(L)}_S = \frac{1}{L}\frac{E^S_{\beta_E}}{\mathcal{V}_{\beta_E}(\tau_{\beta_R}, h_r)}
\end{gather}
was obtained by Wilming \& Gallego~\cite{wilming_gallego} where $T^{(L)}_S$ is the target temperature of the system, $E^S_{\beta_E}~=~\langle H_S \rangle_\beta~=~\text{tr}\{\tau^S_{\beta_E}\ H_S\}$ is the energy of the system in a thermal state at the environmental temperature $\beta_E$ and $\mathcal{V}_{\beta_E}(\tau_{\beta_R}, h_r)$ is a monotone defined in~\cite{wilming_gallego} to determine an agent's ability to cool given a resource $R$. In our case since the resource states are thermal $\mathcal{V}_\beta(\tau_{\beta_R}, h_r)$ simplifies to a difference in free energy between the resource state the a thermal state of the resource at the ambient temperature $\beta_E$.  This gives the expression
\begin{gather}
    T^{(L)}_S = \frac{1}{L}\frac{E^S_{\beta_E}}{\beta_R \Delta F(\tau_{\beta_E}, \tau_{\beta_R})},\label{eq:wilming_gallego}
\end{gather}
where we see that $T^{(L)_S}$ can be made smaller by increasing the number of constituents $L$ of the resource or using colder resources thereby increasing $\Delta F$. Making use of this expression we can examine how the achievable fidelity of the protocol is impacted by Bob having access to more thermal qubits $L$. To do so consider that the fidelity of appending $J$ qubits in the state $\rho_{\ket{0}}(\eta)$ can be expressed as a function of the temperature $T_S$ since $\rho_{\ket{0}}(\eta)$ is a thermal state of $H_S = -\hbar \omega \sigma_z$
\begin{align}
    \mathcal{F} = (1 - \eta)^J = \left(\frac{e^{- \hbar \omega / T_S}}{\mathcal Z}\right)^J = \frac{1}{2^J}\left(1 + \tanh(\hbar \omega/ T_S)\right)^J,
\end{align}
which can be rearranged for $T_S$ recalling that $\text{arctanh}(x) = \frac{1}{2}\ln\left(\frac{1+x}{1-x}\right)$ for $- 1 < x < 1$ we obtain
\begin{align}
T_S &= \frac{2\hbar \omega}{\ln \left(\frac{\mathcal{F}^{1/J}}{1 - \mathcal{F}^{1/J}}\right)} \, : \, 0 < \mathcal{F}^{1/J} < 1.
\end{align}
Recalling eq.\eqref{eq:wilming_gallego} we can use the above relationship to examine how the number of qubits required for cooling $L$ scales with the fidelity Bob obtains in their recovery protocol
\begin{align}
    L &= \frac{E^S_{\beta_E} \ln \left(\frac{\mathcal{F}^{1/J}}{1 - \mathcal{F}^{1/J}}\right)}{2 \hbar \omega \beta_R \Delta F(\tau_{\beta_E},\tau_{\beta_R})},
\end{align}
and so for a fixed number $J$ of qubits to be appended forming the system at a fixed initial state with $\omega$ and $E^S_{\beta_E}$ and fixed resource qubits at a given temperature $\beta_R$ we see the $L$ scales as \begin{align}
    L = \mathcal{O}\left( \ln \left(\frac{\mathcal{F}^{1/J}}{1 - \mathcal{F}^{1/J}}\right)\right)
\end{align}logarithmically with the desired fidelity $\mathcal{F}$ and linearly with the number of qubits to be appended $J$.

\end{document}